\documentclass[twocolumn,superscriptaddress,amssymb,amsmath,nobibnotes,aps,prd,showpacs,nofootinbib]{revtex4-1}
\pdfoutput=1

\usepackage{graphicx,bm,color,psfrag,hyperref}
\usepackage{amsfonts}
\usepackage{lipsum}
\usepackage{caption}
\usepackage{subcaption}
\usepackage{newtxtext}
\usepackage{threeparttable,booktabs}

\newcommand{\be}{\begin{equation}}
\newcommand{\ee}{\end{equation}}
\begin{document}

\title{Cosmic acceleration in a dust only universe via energy-momentum powered gravity}

\author{\"{O}zg\"{u}r Akarsu}
\email{akarsuo@itu.edu.tr}

\author{Nihan Kat{\i}rc{\i}}
\email{nihan.katirci@itu.edu.tr}
\affiliation{Department of Physics, \. Istanbul Technical University, Maslak 34469 \. Istanbul, Turkey}

\author{Suresh Kumar}
\email{suresh.kumar@pilani.bits-pilani.ac.in}
\affiliation{Department of Mathematics, BITS Pilani, Pilani Campus, Rajasthan-333031, India}

%\pacs{}
%%%%%%%%%%%%%%
\begin{abstract}
We propose a modified theory of gravitation constructed by the addition of the term $f(T_{\mu\nu}T^{\mu\nu})$ to the Einstein-Hilbert action, and elaborate a particular case $f(T_{\mu\nu}T^{\mu\nu})=\alpha(T_{\mu\nu}T^{\mu\nu})^{\eta}$, where $\alpha$ and $\eta$ are real constants, dubbed as energy-momentum powered gravity (EMPG). We search for viable cosmologies arising from EMPG especially in the context of the late-time accelerated expansion of the Universe. We investigate the ranges of the EMPG parameters $(\alpha,\eta)$ on theoretical as well as observational grounds leading to the late-time acceleration of the Universe with pressureless matter only, while keeping the successes of standard general relativity at early times. We find that $\eta=0$ corresponds to the $\Lambda$CDM model, whereas $\eta\neq 0$ leads to a $w$CDM-type model. However, the underlying physics of the EMPG model is entirely different in the sense that the energy in the EMPG Universe is sourced by pressureless matter only. Moreover, the energy of the pressureless matter is not conserved, namely, in general it does not dilute as $\rho\propto a^{-3}$ with the expansion of the Universe. Finally, we constrain the parameters of an EMPG-based cosmology with a recent compilation of 28 Hubble parameter measurements, and find that this model describes an evolution of the Universe similar to that in the $\Lambda$CDM model. We briefly discuss that EMPG can be unified with Starobinsky gravity to describe the complete history of the Universe including the inflationary era.

 \end{abstract}
\date{\today}

%\keywords{}
\maketitle
\section{Introduction}

Inflation is a theory of exponential expansion of space in the early Universe corresponding to energy scales $\sim10^{16}\,{\rm GeV}$ \cite{Starobinsky:1980te,Guth:1980zm,Linde:1981mu,Albrecht:1982wi,Baumann:2009ds,Linde:2014nna,Ade:2015lrj}. Besides, today it is confirmed by means of several independent observations that the Universe again started to expand at an accelerated rate approximately $5$ Gyr ago, and it has happened at low energy scales $\sim 10^{-4}\, {\rm eV}$ \cite{Komatsu:2010fb,Aubourg:2014yra,Ade:2015xua}. The totally different energy scales show that the physics behind these two phenomena should be different. Inflation requires a modification of gravitation that plays a role at high energy densities such as the Starobinsky model \cite{Starobinsky:1980te}, while the late-time acceleration of the Universe would be related with a similar mechanism but working at very low energy scales or a modification to the general theory of relativity (GR) that plays a role at sufficiently low energies. The approaches accommodating the current accelerated expansion of the Universe are classified into two main categories, subject to imminent observational discrimination. First is the introduction of a source in GR with a large negative pressure, which is called ``dark energy'' (DE) and is described most frequently by a scalar field or conventional vacuum energy which is mathematically equivalent to the cosmological constant $\Lambda$  \cite{Sahni:1999gb,Peebles:2002gy,Copeland:2006wr,Bamba:2012cp}. The most successful cosmological model capable of predicting the observed pattern of the expansion of the Universe so far is the six-parameter base $\Lambda$CDM model that is simple and in reasonably good agreement with the currently available high-precision data \cite{Komatsu:2010fb,Aubourg:2014yra,Ade:2015xua}. On the other hand, this model suffers from profound theoretical difficulties, such as the cosmological constant and coincidence problems \cite{Weinberg:1988cp,Sahni:1999gb,Carroll:2000fy,Peebles:2002gy,Sahni:2004ai,Bull:2015stt}. There are also observations suggesting small deviations from $\Lambda$CDM in order to describe the current Universe \cite{Aubourg:2014yra,Ade:2015xua,Zhao:2017cud}. On the other hand, we do not have a promising and concrete fundamental theory giving rise to dark energy models more general than the $\Lambda$CDM model that would account for small deviations from $\Lambda$CDM. The second approach is the modification of gravity at large distances, rather than imposing an unknown kind of source, such as in $f(R)$, scalar-tensor theories, Brans-Dicke theories of gravity, etc. \cite{DeFelice:2010aj,Clifton:2011jh,Capozziello:2011et,Nojiri:2017ncd}. The essence of this approach is to modify the form of the coupling between the source described by the energy-momentum tensor (EMT) $T_{\mu\nu}$ and the spacetime geometry described by the metric tensor $g_{\mu\nu}$. Depending on the modifications in the action, this approach can lead to modifications on either the left- or right-hand side of Einstein's field equations (EFE), $R_{\mu\nu}-\frac{1}{2}g_{\mu\nu}R=\kappa T_{\mu\nu}$, where the terms have the usual meaning. In this approach where one avoids introducing a new kind of source, it may be possible to define the corrections that appear on the left-hand side of the EFE, as in the $f(R)$ theories \cite {Nojiri:2010wj}, as a separate effective source by moving them to the right-hand side of the EFE. However, this might not always be trivial or possible, as in, for example, the generalization of $f(R)$ theories via an explicit coupling of an arbitrary function of the Ricci scalar $R$ with the matter Lagrangian density $\mathcal{L}_{\rm m}$ given in Ref. \cite{Harko:2010mv}. Tests of the modified gravity models including nonminimal coupling between matter and geometry, using direct astronomical and astrophysical observations at the galactic or extragalactic scale, are also possible \cite{Harko:2010vs}. The modifications in the action that lead to the modifications on the left-hand side of the EFE are much more commonly studied in the literature compared to the ones that lead to the modification on the right-hand side of the EFE, namely, in the form of how $T_{\mu\nu}$ appears in the field equations. Of course, it could be possible to find a corresponding modification on the left-hand side of the EFE for a modification appearing on the right-hand side of the EFE though this also might not be trivial or possible. Hence, confining ourselves to the modifications in the action leading to modifications on the left-hand side of the EFE may result in missing a large class of successful modified gravity laws that can be obtained from the EMT-type modifications. 

In the following, we briefly describe the EMT type of modifications considered/studied in the literature. In most of the works, each side $-$ both $R$ and $T$ (the trace of the EMT)$-$ were modified and the resulting cosmological implications were studied with the motivation to explain problems or shortcomings of the standard big bang model and Einstein's standard GR, such as inflation, late-time acceleration etc. $\Lambda(T)$ gravity relates cosmic acceleration via the most general form of the EMT, and yields a relativistically covariant model of interacting DE, based on the principle of least action \cite{Poplawski:2006ey}, where $\Lambda$ in the gravitational Lagrangian is a function of the trace of the EMT. The $f(R,T)$ model was proposed and studied in Ref. \cite{Harko:2011kv} with some specific choices of the function $f(R, T)$. The authors of Ref. \cite{Moraes:2016gpe} discussed a complete cosmological scenario from $f(R,T^{\phi})$ gravity theory proposed in Ref. \cite{Harko:2011kv}. In Refs. \cite{Odintsov:2013iba, Haghani:2013oma, Haghani:2014ina}, more generalized modified theories of gravity with $f(R,T,R_{\mu\nu}T^{\mu\nu})$ gravity were studied. One of us has proposed a Lorentz invariant and covariant generalization of GR in Ref. \cite{Arik:2013sti} considering $f(R,T_{\mu\nu}T^{\mu\nu})$. The last two studies are different in the sense that $T_{\mu\nu}T^{\mu\nu}$- or $R_{\mu\nu}T^{\mu\nu}$-like terms still survive for the $T=0$ case, and do not simply reduce to $f(R)$ theories. For instance, for the electromagnetic field, $f(R,T,R_{\mu\nu}T^{\mu\nu})$ and $f(R,T_{\mu\nu}T^{\mu\nu})$ theories differ,  whereas the predictions of $f(R,T)$ and $f(R)$ theories may be the same. Rather than GR, $f(\mathcal{T})$ gravity$-$ called the teleparallel equivalent of GR$-$ was extended via a nonminimal torsion-matter coupling in the action in Ref. \cite{Harko:2014sja}. Going one step further, the coupling of the torsion scalar with the trace of the EMT was studied in $f(\mathcal{T},T)$ gravity \cite{Harko:2014aja}.

As stated above, the self-contraction of the EMT was first proposed in Ref. \cite{Arik:2013sti}. Here we extend it to a more general form $f(R,T_{\mu\nu}T^{\mu\nu})=\alpha(T_{\mu\nu}T^{\mu\nu})^{\eta}$, whereas in Ref. \cite{Arik:2013sti} $-$particular values of its power$-$ $\eta$ were studied. For instance, it was found that the relation between the Hubble parameter and energy density is of a form familiar from the Cardassian expansion studied in the context of late-time cosmic acceleration, or that $\eta=\frac{1}{2}$ leads to a slight deviation from the standard pressureless matter and radiation behavior, violating energy conservation. The higher-order matter terms are reminiscent of the terms (corrections) that arise naturally in loop quantum gravity \cite{Ashtekar:2006wn,Ashtekar:2011ni}, and those in the brane world models \cite{Brax:2003fv}.  In Ref. \cite{Roshan:2016mbt}, the model was analyzed with the $\eta=1$ case, which corresponds to the EMT squared contribution, dubbed energy-momentum squared gravity (EMSG). For this power of the EMT, the correction terms are important only at sufficiently early times, and therefore this model does not give accelerated expansion without any contribution from other extra fields such as scalar fields which can enter the matter Lagrangian. Since the difference appears in the high energy density regime, the charged black hole solution in energy-momentum squared gravity is different from the standard Reissner-Nordstr\" om spacetime \cite{Roshan:2016mbt}. Here, we present a detailed theoretical and observational analysis of energy-momentum powered gravity (EMPG) provided by the general case $f(R,T_{\mu\nu}T^{\mu\nu})=\alpha(T_{\mu\nu}T^{\mu\nu})^{\eta}$ in the context of the late-time accelerated expansion of the Universe. We investigate the ranges of the EMPG model parameters $(\alpha,\eta)$ for viable cosmologies leading to the late-time acceleration of the Universe with pressureless matter only, while keeping the successes of standard general relativity at early times. We demonstrate that only the dust content is sufficient to explain the observed cosmic acceleration. Moreover, the sequence of the matter-dominated phase, deceleration-acceleration transition, and acceleration is obtained similar to the $\Lambda$CDM model in the presence of dust alone. After the first appearance of our study on the arXiv, the $(T_{\mu\nu}T^{\mu\nu})^{\eta}$-type modification to GR was also studied in Ref. \cite{Board:2017ign} where the authors presented  a range of exact solutions for isotropic universes, and discussed their behaviors with reference to the early- and late-time evolution, accelerated
expansion, and the occurrence or avoidance of singularities.

The paper is structured as follows. In the following section, we present the detailed framework of EMPG. In Sec. III, we demonstrate the viable cosmologies arising from EMPG that lead to the late-time acceleration of the Universe. In Sec. IV, we constrain the EMPG model parameters with a recent compilation of 28 Hubble parameter measurements, and discuss the evolution of the EMPG model in contrast with the $\Lambda$CDM model. In Sec. V, we give our concluding remarks and discuss some future perspectives of the study.

\section{Energy-Momentum Powered Gravity}
We start with the action constructed by the addition of the term $f(T_{\mu\nu}T^{\mu\nu})$  \cite{Arik:2013sti} to the Einstein-Hilbert (EH) action\footnote{According to Lovelock's theorem, the cosmological constant $\Lambda$ arises as a constant of nature, which we have set to zero in Eq. \eqref{action} for the reasons discussed in detail in Sec. \ref{subsec:A}.} as follows:
\begin{align}
S=\int \left(\frac{1}{2\kappa}R+f(T_{\mu\nu}T^{\mu\nu})\right)\sqrt{-g}\,{\rm d}^4x+\int \mathcal{L}_{\rm m}\sqrt{-g}\,{\rm d}^4x,
\label{action}
\end{align}
where $\kappa$ is Newton's constant, $R$ is the Ricci scalar, $g$ is the determinant of the metric, and $\mathcal{L}_{\rm m}$ is the Lagrangian density corresponding to the source that would be described by the energy-momentum tensor $T_{\mu\nu}$. We vary the action with respect to the inverse metric as
  \begin{align}
  \delta S=\int\, {\rm d}^4 x \sqrt{-g} \bigg[\frac{1}{2\kappa}\delta R+\frac{\partial f}{\partial(T_{\mu\nu}T^{\mu\nu})}\frac{\delta(T_{\sigma\epsilon}T^{\sigma\epsilon})}{\delta g^{\mu\nu}}\delta g^{\mu\nu} \nonumber \\
  -\frac{1}{2}g_{\mu\nu}\left(R+f(T_{\sigma\epsilon}T^{\sigma\epsilon})\right)\delta g^{\mu\nu} +\frac{1}{\sqrt{-g}}\frac{\delta(\sqrt{-g}\mathcal{L}_{\rm m})}{\delta g^{\mu\nu}}\bigg],
  \end{align} 
  and, as usual, we define the EMT as
  \begin{align}
  \label{tmunudef}
 T_{\mu\nu}=-\frac{2}{\sqrt{-g}}\frac{\delta(\sqrt{-g}\mathcal{L}_{\rm m})}{\delta g^{\mu\nu}}=g_{\mu\nu}\mathcal{L}_{\rm m}-2\frac{\partial \mathcal{L}_{\rm m}}{\partial g^{\mu\nu}},
 \end{align}
which depends only on the metric tensor components, and not on its derivatives. Consequently, the field equations read as follows:
\begin{align}
G_{\mu\nu} =\kappa T_{\mu\nu}+\kappa\left(fg_{\mu\nu}-2\frac{\partial f}{\partial(T_{\mu\nu}T^{\mu\nu})}\theta_{\mu\nu}\right),
\label{genfieldeq}
\end{align}
where $G_{\mu\nu}=R_{\mu\nu}-\frac{1}{2}Rg_{\mu\nu}$ is the Einstein tensor and $\theta_{\mu\nu}$ is the new tensor defined as
\begin{equation}
\begin{aligned}
\theta_{\mu\nu}&= T^{\sigma\epsilon}\frac{\delta T_{\sigma\epsilon}}{\delta g^{\mu\nu}}+T_{\sigma\epsilon}\frac{\delta T^{\sigma\epsilon}}{\delta g^{\mu\nu}}  \\
&=-2\mathcal{L}_{\rm m}\left(T_{\mu\nu}-\frac{1}{2}g_{\mu\nu}T\right)-TT_{\mu\nu}\\
&\quad\quad+2T_{\mu}^{\gamma}T_{\nu\gamma}-4T^{\sigma\epsilon}\frac{\partial^2 \mathcal{L}_{\rm m}}{\partial g^{\mu\nu} \partial g^{\sigma\epsilon}}.
\label{theta}
\end{aligned}
\end{equation}
We note that the EMT given in Eq. \eqref{tmunudef} does not include the second variation of $\mathcal{L}_{\rm m}$, and hence the last term of Eq. \eqref{theta} is null. As the definition of the matter Lagrangian that gives the perfect-fluid EMT is not unique, one could choose either $\mathcal{L}_{\rm m}=p$ or $\mathcal{L}_{\rm m}=-\rho$, which provide the same EMT (see \cite {Bertolami:2008ab,Faraoni:2009rk} for a detailed discussion). In the present study, we consider $\mathcal{L}_{\rm m}=p$. 

We proceed with a particular form of the model as
\begin{equation}
f(T_{\mu\nu}T^{\mu\nu})=\alpha(T_{\mu\nu}T^{\mu\nu})^{\eta},
\end{equation}
which we shall refer to as the EMPG model. The action now reads as
   \begin{equation}
S=\int \left[\frac{1}{2\kappa}R+\alpha(T_{\mu\nu}T^{\mu\nu})^{\eta}+\mathcal{L}_{\rm m}\right]\sqrt{-g}\,{\rm d}^4x,
\label{eq:action}
\end{equation}
where $\eta$ is the power of the self-contraction of the EMT, and $\alpha$ is a constant that would take part in  determining the coupling strength of the EMT-powered modification to gravity. The Einstein field equations \eqref{genfieldeq} for this action become
  \begin{equation}
G_{\mu\nu}=\kappa T_{\mu\nu}+\kappa \alpha (T_{\sigma\epsilon}T^{\sigma\epsilon})^{\eta}\left[g_{\mu\nu}-2\eta\frac{\theta_{\mu\nu}}{T_{\sigma\epsilon}T^{\sigma\epsilon}}\right].
\label{fieldeq}
\end{equation} 
Using Eq. \eqref{fieldeq}, the covariant divergence of the EMT reads as
\begin{equation}
\begin{aligned}
\label{nonconservedenergy}
\nabla^{\mu}T_{\mu\nu}=&-\alpha g_{\mu\nu}\nabla^{\mu}(T_{\sigma\epsilon}T^{\sigma\epsilon})^{\eta}\\ 
&+2\alpha\eta\nabla^{\mu}\left(\frac{\theta_{\mu\nu}}{(T_{\sigma\epsilon}T^{\sigma\epsilon})^{1-\eta}}\right).
\end{aligned}
\end{equation}
We notice that, in our model, the EMT is not conserved in general since the right-hand side of this equation vanishes only for some particular values of the constants.

In this paper, we shall study the gravity model under consideration in the context of cosmology. Therefore, we proceed by considering the spatially maximally symmetric spacetime metric, i.e., the Robertson-Walker metric, with flat space-like sections
\begin{align}
\label{RW}
ds^2=-dt^2+a^2\,({\rm d}x^2+{\rm d}y^2+{\rm d}z^2),
\end{align}  
where the scale factor $a=a(t)$ is a function of cosmic time $t$ only, and the perfect fluid form of the EMT is given by
\begin{align}
\label{em}
T_{\mu\nu}=(\rho+p)u_{\mu}u_{\nu}+p g_{\mu\nu},
\end{align} 
where $\rho$ is the energy density, $p$ is the thermodynamic pressure, and $u_{\mu}$ is the four-velocity satisfying the conditions $u_{\mu}u^{\mu}=-1$ and $\nabla_{\nu}u^{\mu}u_{\mu}=0$. We assume a barotropic equation-of-state (EoS),
\begin{align}
\label{eos}
\frac{p}{\rho}=w={\rm const.}
\end{align}
to describe the physical ingredient of the Universe. Using Eqs. \eqref{em} and \eqref{eos}, we find $\theta_{\mu\nu}$ given in Eq. \eqref{theta} and the self-contraction of the EMT as 
\begin{align}
\label{thetafrw}
\theta_{\mu\nu}&=-\rho^2(3w+1)(w+1)u_{\mu}u_{\nu},\\
 T_{\mu\nu}T^{\mu\nu}&=\rho^2(3w^2+1),
\label{trace}
\end{align}
respectively. Then, using Eqs. \eqref{thetafrw} and \eqref{trace} as well as the metric \eqref{RW} in the field equations \eqref{fieldeq}, we obtain the following set of two linearly independent differential equations in two unknown functions $H$ and $\rho$:
%\begin{align}
%3 H^2=\kappa\rho+\kappa\alpha \rho^{2\eta}(3w^2+1)^{\eta}\left[2\eta-1+\frac{8w\eta}{3w^2+1}\right], \label{eq:rho}\\
%-2\dot{H}-3H^2=\kappa w\rho+\kappa \alpha \rho^{2\eta}(3w^2+1)^{\eta}, \label{eq:pres}
%\end{align}
\begin{align}
3 H^2=&\kappa\rho+\kappa'  \rho_0\left(\frac{\rho}{\rho_0}\right)^{2\eta} , \label{eq:rhoprime}\\
-2\dot{H}-3H^2=&\kappa w\rho+ \frac{\kappa' \rho_0}{2\eta-1+\frac{8w\eta}{3w^2+1}} \left(\frac{\rho}{\rho_0}\right)^{2\eta}, \label{eq:presprime}
\end{align}
where $H=\dot{a}/a$ is the Hubble parameter and the subscript $0$ refers to the present-day values of the parameters.
%. Using the time-redshift transfomation $\frac{{\rm d}z}{{\rm d}t}=-H(1+z)$, we recast the field equations \eqref{eq:rho}-\eqref{eq:pres} as follows:
%\begin{align}
%3 H^2=&\kappa\rho+\kappa'  \rho_0\left(\frac{\rho}{\rho_0}\right)^{2\eta}, \label{eq:rhoprime}\\
%2HH'\,(1+z)-3H^2=&\kappa w\rho+ \frac{\kappa' \rho_0}{2\eta-1+\frac{8w\eta}{3w^2+1}} \left(\frac{\rho}{\rho_0}\right)^{2\eta}, 
%\label{eq:presprime}
%\end{align}
%where prime is the derivative with respect to $z$
The constant $\kappa'$ is the gravitational coupling of the EMT-powered modification, and is given by
\begin{equation}
\kappa'=\alpha' \kappa= \alpha\,\kappa \rho_{0}^{2\eta-1}(3w^2+1)^{\eta}\left[2\eta-1 +\frac{8w\eta}{3w^2+1}\right].
\label{eq:alphaprime}
\end{equation}

We note that the source (i.e., $T^{\mu\nu}$) that we considered in the EMT-powered term is the same as the one obtained from $\mathcal{L}_{\rm m}$, but the terms that appear in the field equations due to the EMT-powered term in the action couple to gravity with a different strength as $\kappa'=\alpha'\kappa$, where $\alpha'$ is the ratio of this coupling with respect to the conventional Newtonian coupling $\kappa$. We further note that $\alpha'=\alpha'(\alpha,\eta,\rho_{0},w)$, namely, $\alpha'$ depends not only on $\alpha$ but also on the current energy density $\rho_0$, and the type of source described by EoS parameter $w$ provided that $\eta\neq0$. This implies that the EMT-powered term would lead to a violation of the equivalence principle, which is intimately connected with some of the basic aspects of the unification of gravity with particle physics such as string theories (see Ref. \cite{Uzan:2010pm} and references therein). We will not elaborate the implications of this property of our model in this paper, since here we shall discuss the dynamics of the Universe in the presence of only a pressureless fluid, i.e., a monofluid Universe, with the purpose of describing the late-time acceleration of the Universe (where the radiation is negligible) without invoking a cosmological constant or any dark energy source.

Next, we note that the first Friedmann equation \eqref{eq:rhoprime} is in the form of the well-known Cardassian expansion  ($H^2=A\rho+B\rho^n$, with $A$, $B$, and $n$ being constants) \cite{Freese:2002sq}, which was motivated by the fact that the term of the form $\rho^n$ can generically appear as a consequence of embedding the observable Universe as a brane in extra dimensions \cite{Freese:2002sq,Chung:1999zs}. On the other hand, in the second Friedmann equation \eqref{eq:presprime}, we see that the additional pressure term (the latter term that appears due to the EMPG) is in the form of the pressure of the generalized Chaplygin gas ($p=-A/\rho^{\alpha}$, where $A$ is a positive constant) \cite{Kamenshchik:2001cp,Bento:2002ps}. Also, for the special case $\eta=1$, the total pressure [the whole right-hand side of Eq. \eqref{eq:presprime}] is similar to the quadratic equation of state ($p=p_0+\alpha\rho+\beta\rho^2$, where $p_0$, $\alpha$, and $\beta$ are constants) of dark energy \cite{Ananda:2005xp}. However, one may check that our model in fact does not correspond to any of them, namely, the modified Friedmann equations of our model \eqref{eq:rhoprime} and \eqref{eq:presprime} do not simultaneously match the Friedmann equations of each of these models. The main reason behind this is the violation of the local/covariant energy-momentum conservation in our model. The corresponding energy conservation equation \eqref{nonconservedenergy} is
\begin{align}
\label{noncons}
&\dot \rho+ 3H(1+w)\rho=\nonumber \\
&-2\alpha'\eta\, \left[ \dot \rho +3H \rho\left(\frac{1+\frac{4w}{3w^2+1}}{2\eta-1+\frac{8w\eta}{3w^2+1}}\right) \right]\,\left(\frac{\rho}{\rho_0}\right)^{2\eta-1}.
\end{align}
Here we can see that the local/covariant energy-momentum conservation $\nabla^{\mu}T_{\mu\nu}=0$, which would lead to $\rho\propto a^{-3(1+w)}$, is not satisfied for $\alpha'\neq0$ in general. Some particular cases, in which the right-hand side of the equation vanishes are as follows: (i) the case $\eta=0$, which is trivial; (ii) the case $w=-1$, i.e., the conventional vacuum energy; and (iii) the case $\eta=\frac{3w^2+3w+2}{2(3w+1)(w+1)}$, which gives $\eta=1$ for $w=0$ (dust) and $\eta=\frac{5}{8}$ for $w=\frac{1}{3}$ (radiation). One may also check that the standard energy conservation is not satisfied for the range $\frac{24-9\sqrt{6}}{48-20\sqrt{6}}<\eta<\frac{1}{2}$. In the context of late-time acceleration, the violation of energy conservation is not uncommon in the literature \cite{Josset:2016vrq,Shabani:2017vns}.

\section{Late-time acceleration in a dust-only Universe}

The late-time acceleration of the Universe takes place at relatively low energies, and hence it would be wise to search for suitable ranges of the model parameters $\alpha'$ and $\eta$ in which our modification is effective at sufficiently low energy densities, but negligible at high energy densities, namely, at energies higher than that of recombination. In this way, the successes of the standard cosmology would be untouched, and we would be able to derive accelerated expansion. Knowing that the energy density of the matter source $\rho$ should be positive, we would like to ensure that the EMT-powered contribution to the Hubble parameter [the latter term in Eq. \eqref{eq:rhoprime}] is positive as well. So we set $\kappa'>0$, implying  that $\alpha'>0$. In the $\alpha'=0$ case, the EMT-powered modification vanishes, and our model reduces to the standard Friedmann model as the standard EH action is recovered. In this case, we get a matter (e.g., dust) dominated Universe in GR, and to get accelerated expansion of the Universe we need a cosmological constant or DE source. Therefore, in what follows, we consider $\alpha'>0$, focus our attention on the free parameter $\eta$ and investigate the model.

\subsection{$\Lambda$CDM-type behavior ($\eta=0$) without $\Lambda$}
\label{subsec:A}

Let us start with the special case $\eta=0$. In this case, in the presence of pressureless matter $\rho_{\rm m}$ with $w=w_{\rm m}=0$, from Eqs. \eqref{eq:rhoprime} and \eqref{noncons} we obtain $3H^2=\kappa\rho_{{\rm m},0}\left( \frac{a}{a_0}\right)^{-3}+\kappa' \rho_{{\rm m},0}$. Hence, the model yields the same mathematical structure as the $\Lambda$CDM model  $3H^2=\kappa\rho_{{\rm m},0}\left( \frac{a}{a_0}\right)^{-3}+\kappa\rho_{\Lambda}$, where $\rho_{\Lambda}=\rm const$. But we stress that the underlying physics of these two models are completely different. Namely, the only source in our model is pressureless matter, which is not the case in the $\Lambda$CDM model. For instance, if we consider an empty Universe ($\rho_{{\rm m},0}=0$), we find $H=0$ (a static Universe) in our model, whereas we find $H=\sqrt{\frac{\rho_{\Lambda}}{3}}=\rm const$ (the de Sitter solution) in the $\Lambda$CDM model. On the other hand, we can make use of this correspondence to estimate the value of $\alpha'$. Setting $a=a_0$ in both models, we obtain $3H_0^2=(\kappa+\kappa')\rho_{{\rm m},0}$ for our model, and $3H_0^2=\kappa(\rho_{{\rm m},0}+\rho_{\Lambda})$ for the $\Lambda$CDM model. Then, using these we obtain the following correspondence between the parameters of these two models: $\alpha'=\rho_{\Lambda}/\rho_{{\rm m},0}$. Hence, from the recent Planck results \cite{Ade:2015xua} giving $\Omega_{\Lambda,0}=\frac{\rho_{\Lambda}}{\rho_{{\rm m},0}+\rho_{\Lambda}}\sim 0.69$ for the current Universe, we estimate that $\alpha' \sim 2.2$\,. Knowing that $\Lambda$CDM is very successful in describing the observed Universe, it would not be wrong to conclude from this simple investigation that our model with $\eta\sim0$ and $\alpha'\sim2$ would successfully describe the background dynamics of the observed Universe. We note that the role of the cosmological constant in $\Lambda$CDM has been taken over by the energy density of the pressureless matter itself due to its new form of coupling to gravity which is about $2$ times stronger than that of the conventional coupling. On the other hand, we could also obtain accelerated expansion by including a cosmological constant in our model from the beginning (which is obvious), or by considering vacuum energy for the EMT. In the latter case$-$namely, when we consider the conventional vacuum energy described by the EoS parameter $w=w_{\rm vac}=-1$$-$irrespective of the value of $\eta$, our model gives de sitter expansion, as in standard GR, but with an enhanced Hubble parameter $H=\sqrt{\frac{1}{3}\rho_{\rm vac,0}(\kappa+\kappa')}={\rm const.}$ due to the coupling $\kappa'$ of the EMT-powered modification. However, obtaining accelerated expansion without invoking a cosmological constant or vacuum energy $-$ in contrast to $\Lambda$CDM $-$ would render our model immune to the well-known cosmological constant problem, namely, the extreme fine-tuning of the value of $\Lambda$. Although we know of no special symmetry that could enforce a vanishing vacuum energy while remaining consistent with the known laws of physics, it is usually thought to be easier to imagine an unknown mechanism that would set $\Lambda$ precisely to zero than one that would suppress it by just the right amount $\rho_{\Lambda}^{\rm (observation)}/\rho_{\Lambda}^{(\rm theory)}\sim10^{-120}$ to yield an observationally accessible cosmological constant (see Refs. \cite {Weinberg:1988cp,Sahni:1999gb,Carroll:2000fy,Peebles:2002gy,Sahni:2004ai,Bull:2015stt} and references therein). In our model, on the other hand, in the presence of only a pressureless fluid the observations would favor $\kappa' \sim 2\kappa$, which means that the coupling strength of the corresponding energy density of the EMT-powered modification to gravity has the same order of magnitude as the conventional coupling of the energy density to gravity.

It may be noted that we have avoided the introduction of a nonzero cosmological constant in the action \eqref{action}, though according to Lovelock's theorem\footnote{Lovelock's theorem \cite{Lovelock:1971yv,Lovelock:1972vz} states that the only possible second-order Euler-Lagrange expression obtainable in a four-dimensional space from a scalar density of the form $\mathcal{L}= \mathcal{L}(g_{\mu\nu})$ is $E_{\mu\nu}=\sqrt{-g}\left(\lambda_1 G_{\mu\nu}+\lambda_2 g_{\mu\nu}\right)$, where $\lambda_1$ and $\lambda_2$ are constants, leading to Newton's gravitational constant $G$ and cosmological constant $\Lambda$ in Einstein's field equations $G_{\mu\nu}+\Lambda g_{\mu\nu}=\kappa T_{\mu\nu}$ (see \cite{Bull:2015stt,Clifton:2011jh,Straumann} for further reading).} the cosmological constant $\Lambda$ arises as a constant of nature like Newton's gravitational constant $G=\frac{\kappa}{8\pi}$. Indeed, if we stick to the usual interpretation of $\Lambda$ as the term representing the vacuum energy $\rho_{\rm vac}$, one then gets $R=4\Lambda=4\kappa \rho_{\rm vac}$ in GR. However, then the fine-tuning problem arises, as we discussed above, as well as some other problems as well inherited from the standard $\Lambda$CDM model due to the presence of $\Lambda$ (see Ref. \cite{Bull:2015stt} for a recent review). Similarly, in our model, as mentioned above, in the presence of only pure vacuum $w_{\rm vac}=-1$, we obtain $R=4({\kappa+\kappa'}) \rho_{\rm vac}$. Therefore, considering a dust-only ($w=0$) Universe implies that a constant of nature, which would correspond to $\Lambda=({\kappa+\kappa'}) \rho_{\rm vac}$, is set to zero in the construction of our model. On the other hand, as discussed in the previous paragraph, setting $\Lambda$ to zero (i.e., considering only dust without vacuum energy) alleviates the fine-tuning problem in our model. However, there are several other reasons for such a setting within the scope of the current study. If we consider a nonzero $\Lambda$, then we can see from the field equations \eqref{eq:rhoprime} and\eqref{eq:presprime} that, for the case $\eta=0$, the constant contribution from dust itself, i.e., $\kappa'\rho_0$, and $\Lambda$ would be degenerate at least at the background level. In addition, it is conceivable that such a degeneracy would also appear for the case $\eta\sim0$. Besides, for the case $\eta>\frac{1}{2}$, the new terms that arise due to EMPG with respect to GR would be suppressed at relatively low energy densities, and then one can recover the $\Lambda$CDM model in the late Universe and let EMPG alter the early Universe. For instance, in \cite{Roshan:2016mbt} the Universe filled with dust approached the $\Lambda$CDM model at late times in EMSG, i.e., the special case ${\rm EMPG}_{\eta=1}$ of our model. Indeed, in Ref. \cite{ACEK}, the parameter $\alpha$ was well constrained from neutron stars and it was found that EMSG can lead to a significant modification in the dynamics of the Universe only when the age of the Universe is $t\lesssim10^{-4}\,{\rm s}$, long before the physical processes relevant to big bang nucleosynthesis that take place when $t\sim 10^{-2}\,-10^{2}\,{\rm s}$, and leaves the standard cosmology unaltered \cite{ACEK}. It is clear from these studies that our model in the presence of $\Lambda$ would create a degeneracy between the cases $\eta\sim0$ and $\eta>\frac{1}{2}$ when subjected to the data from cosmological observations. Moreover, when we carry out our observational analysis, the case with $\eta\sim0$, $\alpha'\sim 2.2$, and $\Lambda\sim0$ would be degenerate with the case $\eta>\frac{1}{2}$, $\frac{\rho_{{\rm m},0}}{\rho_{\Lambda}+\rho_{{\rm m},0}}\sim 0.7$ for a range of $\alpha'$ depending on the value of $\eta$. However, it is possible to overcome some of the above-mentioned degeneracies, e.g., by considering compact astrophysical objects such as neutron stars for the case $\eta>\frac{1}{2}$ (e.g., in Ref. \cite{ACEK} EMSG (the case ${\rm EMPG}_{\eta=1}$) was studied) , although such an analysis is beyond the scope of the current study. Finally, we would like to mention that the inclusion of $\Lambda$ in the action \eqref{action} may provide a much richer theory and we postpone this idea to future work.

\subsection{Viable cosmologies more general than $\Lambda$CDM without $\Lambda$}

In the above, we have demonstrated that our model for the case $\eta=0$ provides not only the same background dynamics as the $\Lambda$CDM model but also some additional promising features. Let us now elaborate our discussion by considering the $\eta\neq0$ cases. We first explore the range for the value of $\eta$ to get accelerated expansion in the relatively late Universe. We note from Eq. \eqref{eq:rhoprime} that the parameter $\eta$ determines whether the EMPG modification would come into play at large or small values of $\rho$. In the case $\eta=\frac{1}{2}$, the terms from the standard EH and the EMT-powered modification in Eq. \eqref{eq:rhoprime} have the same power, and they track each other, i.e., their relative contributions to $H$ do not depend on the value of the energy density, but only on $\alpha'$ only. In the case of $\eta>\frac{1}{2}$, the EMT-powered term manifests itself at  larger values of $\rho$, namely, in the relatively early Universe, say, at a time before big bang nucleosynthesis (BBN) processes took place; in this case, one could discuss alternative scenarios for the beginning of Universe. For instance, in the model introduced in Ref. \cite{Roshan:2016mbt}, which corresponds to the case $\eta=1$ of our model in the presence of $\Lambda$, the Universe may not reach an initial singularity and bounce when $\rho$ increases to a certain value in the early Universe, and at sufficiently low values of $\rho$ the model becomes indistinguishable from GR and hence it is $\Lambda$ that leads to a late-time accelerated expansion. In the case $\eta<\frac{1}{2}$, the EMT$-$powered term manifests itself at lower values of $\rho$ (say, in the present Universe $\frac{\rho_{\rm m}}{\rho_{{\rm m},0}} \sim 1$), where it may play a role in the accelerated expansion of the Universe, thereby providing an alternative to $\Lambda$ or dark energy sources while keeping the successes of the standard cosmology based on GR at earlier times of the Universe. For instance, the particular case $\eta=\frac{1}{4}$ of our model may lead to a late-time accelerated expansion as discussed in Ref. \cite{Arik:2013sti}.

\subsubsection{Accelerated expansion at low energies}

In light of the above discussion, we see that we should consider the case $\alpha'>0$ with $\eta<\frac{1}{2}$ for the EMT-powered term to be effective at lower values of the energy density. As a further step, we write the deceleration parameter $q=-1-\frac{\dot{H}}{H^2}$ to check whether the EMT-powered modification would give rise to the accelerated expansion,
\begin{align}
\label{decpar}
q=\frac{1+3w+\alpha'\left(\frac{\rho}{\rho_0}\right)^{2\eta-1}\left[1+\frac{3}{2\eta-1+\frac{8w\eta}{3w^2+1}}\right]}
{2+2\alpha'\left(\frac{\rho}{\rho_0}\right)^{2\eta-1}}.
\end{align}
As we discussed above, the EMPG modification becomes negligible at larger values of $\rho$ for $\eta<\frac{1}{2}$. It is possible to set the values of $\alpha'$ and $\eta$ such that at energy scales larger than that of recombination, our model would be indistinguishable from GR, i.e., the terms with $\alpha'$ in Eqs. \eqref{decpar} and \eqref{noncons} would be negligible and lead to $q=\frac{1+3w}{2}$ and $\dot \rho+ 3H(1+w)\rho=0$, respectively. Hence, the standard cosmology would be left unaltered for times before recombination that took place at $z\approx1100$. Because the Universe should always be matter dominated (hence $w=0$, implying $q=\frac{1}{2}$) at recombination \cite{Barrow:1976rda} and we introduce no sources (such as $\Lambda$ or DE) that would dominate over pressureless matter after recombination, we can assume that the Universe is only filled with pressureless matter $\rho_{\rm m}$. Hence, we use the deceleration parameter 
\begin{align}
\label{qdef}
q=\frac{1+2\alpha'\left(\frac{\eta+1}{2\eta-1}\right)\left(\frac{\rho_{\rm m}}{\rho_{{\rm m},0}}\right)^{2\eta-1}}
{2+2\alpha'\left(\frac{\rho_{\rm m}}{\rho_{{\rm m},0}}\right)^{2\eta-1}},
\end{align} 
obtained by substituting $w=0$ in Eq. \eqref{decpar}, to investigate the evolution the Universe at $z\lesssim1100$ as well as the late-time accelerated expansion, which starts at $z\sim 0.6$. Because the condition $\alpha'>0$ makes the denominator always positive, $q$ can take negative values if
\begin{equation}
\label{cond1}
-1<\eta<\frac{1}{2},
\end{equation}
due to the term $\frac{\eta+1}{2\eta-1}$ in the numerator. We note that the upper limit coincides with the one we obtained in order for the EMPG modification to be effective at lower energy density values. Thus, the condition \eqref{cond1} guarantees standard GR at earlier times, and cosmic acceleration at later times by means of the EMPG modification. Finally, we get accelerated expansion in the present Universe at $z=0$, implying $\rho_{\rm m}=\rho_{{\rm m},0}$, provided that
\begin{equation}
\label{cond2}
-1+\frac{3}{2\alpha'+2}<\eta<\frac{1}{2},
\end{equation}
which is a stronger condition than the one given in Eq. \eqref{cond1} for all positive values of $\alpha'$. We next see from Eq. \eqref{qdef} that
\begin{equation}
q\approx\frac{\eta+1}{2\eta-1}\quad\textnormal{for}\quad \rho_{\rm m}\ll\rho_{{\rm m},0},
\end{equation}
provided that $\rho_{\rm m}$ can decrease to sufficiently small values under the condition \eqref{cond1}. (We shall discuss the minimum $\rho_{\rm m}$ values that can be achieved at the end of this section.) According to this, the Universe evolves toward $-1<q<0$ for $-1<\eta<0$, toward $q=-1$ for $\eta=0$, and toward $q<-1$ for $0<\eta<\frac{1}{2}$. These are in line with the effective energy density and pressure $\rho'= \rho_{{\rm m},0}\left(\frac{\rho_{\rm m}}{\rho_{{\rm m},0}}\right)^{2\eta}$ and $p'= \frac{\rho_{{\rm m},0}}{2\eta-1}\left(\frac{\rho_{\rm m}}{\rho_{{\rm m},0}}\right)^{2\eta}$ from the EMPG modification in the modified Friedmann equations \eqref{eq:rhoprime}$-$\eqref{eq:presprime}, respectively. In the presence of only pressureless matter, we find that the effective EoS parameter of the EMPG modification is $w'=\frac{p'}{\rho'}=\frac{1}{2\eta-1}$, which gives a quintessence-like EoS parameter  $-1<w'<-\frac{1}{3}$ for $-1<\eta<0$, a vacuum-like EoS parameter $w'=-1$ for $\eta=0$, and a phantom-like EoS parameter $w'<-1$ for $0<\eta<\frac{1}{2}$. These signal that our model would lead to a Universe that exhibits a similar (but not the same, since covariant energy-momentum conservation is not always satisfied in our model) evolution as that of the well-known $w$CDM model, which is the simplest (phenomenological) and most widely considered extension of the $\Lambda$CDM model, in observational cosmology. Based on GR, in the $w$CDM model (where $w$ represents the EoS of the DE source as $w_{\rm DE}={\rm constant}$)  the energy densities of pressureless matter and DE evolve as $\rho_{\rm m} \propto a^{-3}$ and $\rho_{\rm DE} \propto a^{-3(1+w_{\rm DE})}$, respectively, due to the covariant conservation of the EMT tensors of these sources. These imply that $\rho_{\rm DE} \propto \rho_{\rm m}^{(1+w_{\rm DE})}$, which leads to the following deceleration parameter for the spatially flat $w$CDM model:
 \begin{align}
\label{qdefw1}
q_{w{\rm CDM}}=\frac{1+(3{w_{\rm DE}+1)\,\frac{1-\Omega_{{\rm m},0}}{\Omega_{{\rm m},0}}\left(\frac{\rho_{\rm m}}{\rho_{{\rm m},0}}\right)^{w_{\rm DE}}}}{2+2\,\frac{1-\Omega_{{\rm m},0}}{\Omega_{{\rm m},0}}\left(\frac{\rho_{\rm m}}{\rho_{{\rm m},0}}\right)^{w_{\rm DE}}}.
\end{align}
Its mathematical form is very similar to our model's deceleration parameter \eqref{qdef}. Indeed, under the  transformations $w_{\rm DE} \rightarrow 2\eta-1$ and $\frac{1-\Omega_{{\rm m},0}}{\Omega_{{\rm m},0}}\rightarrow\alpha'$, this can be recast as follows:
\begin{align}
\label{qdefw2}
q_{w{\rm CDM}}=\frac{1+2\alpha' (3\eta-1)\left(\frac{\rho_{\rm m}}{\rho_{{\rm m},0}}\right)^{2\eta-1}}
{2+2\alpha'\left(\frac{\rho_{\rm m}}{\rho_{{\rm m},0}}\right)^{2\eta-1}}.
\end{align}
However, we note that the coefficients in front of the latter terms in the numerators in Eqs. \eqref{qdef} and \eqref{qdefw2} are not the same, except for the cases $\eta=0$ corresponding to $w'=w_{\rm DE}=-1$ ($\Lambda$/conventional vacuum) and $\eta=1$ corresponding to $w'=w_{\rm DE}=1$, i.e., a stiff fluid that does not lead to accelerated expansion and hence cannot be a DE candidate. 
\begin{figure}[t!!]
\captionsetup{justification=raggedright,singlelinecheck=false,font=footnotesize}
\includegraphics[width=0.5\textwidth]{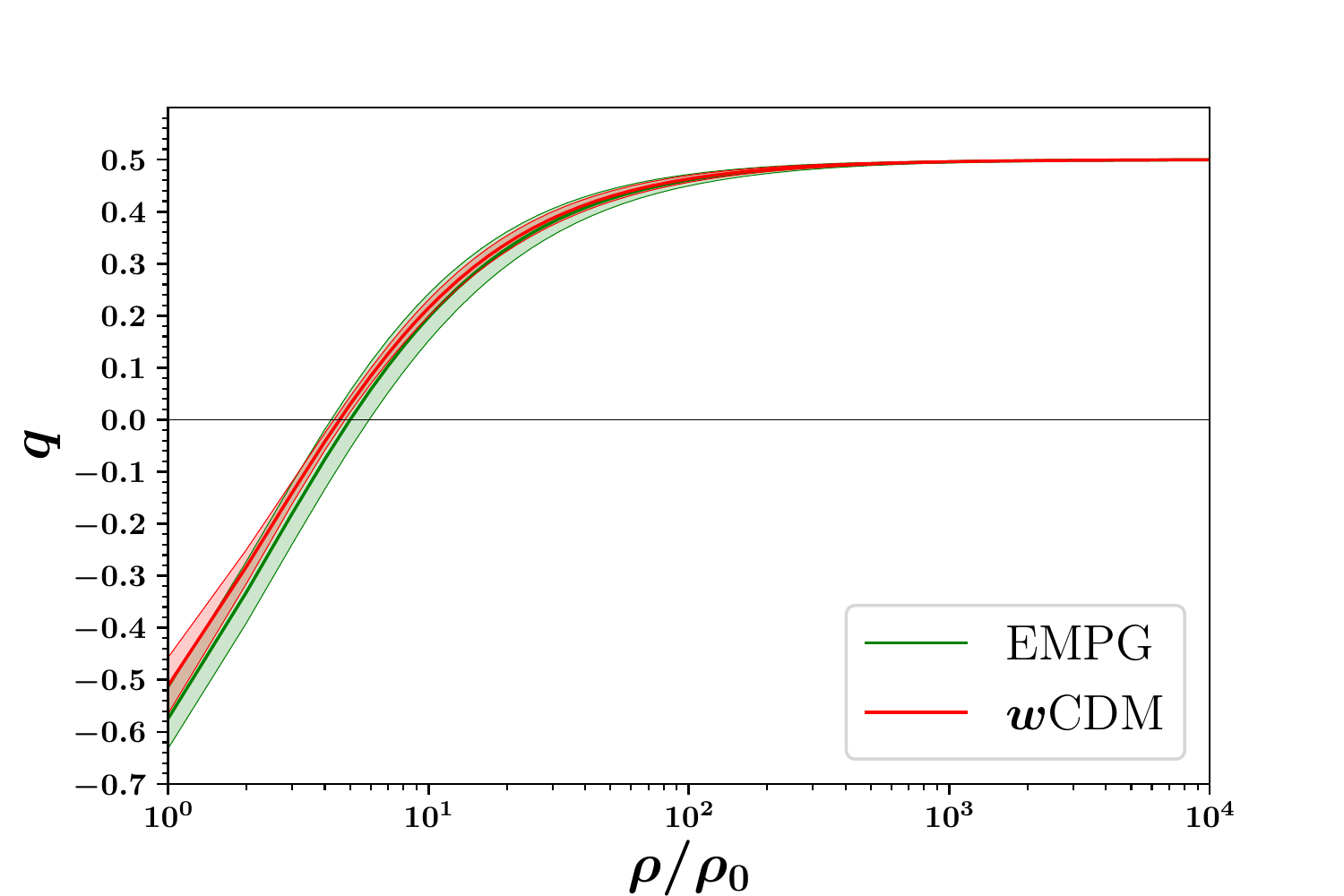}
\caption{$q$ versus $\frac{\rho_{\rm m}}{\rho_{{\rm m},0}}$ curve of the $w$CDM and EMPG models. The figure is plotted by using $\Omega_{{\rm m},0}=0.305\pm 0.010$ and $w_{\rm DE}=-0.97\pm0.05$, corresponding to $\alpha'=2.28\pm0.11$ and $\eta=0.015\pm0.025$ according to the transformations between the two models, from the constraints for the $w$CDM model from Planck+BAO+SN data presented in Ref. \cite{Aubourg:2014yra}.} \label{qvsrho}
\end{figure}
In Fig. \ref{qvsrho}, we show that the EMPG [Eq. \eqref{qdef}] and $w$CDM [ Eq. \eqref{qdefw1}] models differ only slightly by depicting the deceleration parameter versus the energy density of the pressureless matter, using $\Omega_{{\rm m},0}=0.305\pm 0.010$ and $w_{\rm DE}=-0.97\pm0.05$, corresponding to $\alpha'=2.28\pm0.11$ and $\eta=0.015\pm0.025$ according to the above transformations, from the constraints for the $w$CDM model from Planck+BAO+SN data presented in Ref. \cite{Aubourg:2014yra}. We see that our model would lead to exactly the same behavior as in $\Lambda$CDM model when $\eta=0$, though the physics is different, and would also lead to a behavior similar to that in the $w$CDM model when $\eta\neq0$. The DE source of the $w$CDM model is phenomenological rather than derived from a fundamental theory, and may be described by a scalar field yielding a suitable potential, e.g., as it was reconstructed analytically to give $\rho_{\rm DE}\propto a^{-3(1+w_{\rm DE})}$ in two fluid cosmological models in Ref. \cite{Rubano:2001xi}. The case $w_{\rm DE}<-1$ for the $w$CDM model corresponds to phantom DE, which would be described by a noncanonical scalar field with a negative kinetic term, which would eventually lead to $q<-1$, i.e., the Universe would end in a big rip \cite{Caldwell:1999ew,Caldwell:2003vq}. On the other hand, the behavior we would get in the range $0<\eta<\frac{1}{2}$ in the presence of ordinary matter only does the job of the scalar field with a negative kinetic term, namely, the phantom field. We know that models with a negative kinetic term suffer from the problem of quantum instability due to an unbounded vacuum state from below in the matter sector, and hence they are problematic \cite{carroll03,cline04}. It is interesting that we obtained a $w$CDM-like behavior without invoking a scalar field with a particular kind of potential that is not from a fundamental theory, but rather reconstructed and requires a negative kinetic term for the case $w_{\rm DE}<-1$.

\subsubsection{Evolution of the energy density}

The energy conservation equation \eqref{noncons} for $\rho=\rho_{\rm m}$ with $w=0$ reads as
\begin{align}
\label{nonconser}
&\dot \rho_{\rm m}+ 3H\rho_{\rm m}=-2\alpha'\eta\, \left[ \dot \rho_{\rm m} +\frac{3H \rho_{\rm m}}{2\eta-1} \right]\,\left(\frac{\rho_{\rm m}}{\rho_{{\rm m},0}}\right)^{2\eta-1},
\end{align}
and it gives the time rate of change of the matter energy density as
\begin{align}
\label{eqn:conh}
\frac{\dot{\rho}_{\rm m}}{\rho_{\rm m}} = -3 H \, \frac{1+\frac{2\alpha'\eta}{2\eta-1}\left(\frac{\rho_{\rm m}}{\rho_{{\rm m},0}}\right)^{2\eta-1}}{1+2\alpha'\eta \left(\frac{\rho_{\rm m}}{\rho_{{\rm m},0}}\right)^{2\eta-1}}.
\end{align}
Taking the derivative of Eq. \eqref{eq:rhoprime} with respect to time $t$, and then using the result together with Eq. \eqref{eqn:conh}, we find
\begin{equation}
\label{eqn:hdot}
\dot{H}=-\frac{\kappa}{2}\rho_{\rm m} \left[1+\frac{2\alpha' \eta}{2\eta-1} \left(\frac{\rho_{\rm m}}{\rho_{{\rm m},0}}\right)^{2\eta-1}   \right].
\end{equation}
Among many other features that are required for a successful cosmological model, a viable cosmological model would be expected to satisfy the conditions $H>0$, $\dot{H}<0$, $\rho_{\rm m}>0$, and $\dot \rho_{\rm m}<0$ when $\rho_{\rm m}>\rho_{{\rm m},0}$, at least since after the matter-radiation equality that took place about $z\sim3600$ until the present time $z\sim0$ ($\rho_{\rm m}\sim\rho_{{\rm m},0}$), in line with the standard cosmological model. Accordingly, using Eqs. \eqref{eq:rhoprime}, \eqref{eqn:conh}, and \eqref{eqn:hdot}, we find that these conditions imply the following relations between $\alpha'$ and $\eta$:
\begin{equation}
\begin{aligned}
\label{eqn:fcon}
\alpha'<-\frac{1}{2\eta}&\quad\textnormal{for}\quad -1<\eta<0,\\
\alpha'<-1+\frac{1}{\eta}&\quad \textnormal{for}\quad 0<\eta<\frac{1}{2},
\end{aligned}
\end{equation}
These are in addition to the condition \eqref{cond1} that guarantees that EMPG gravity approaches GR at high energy densities, and leads to accelerated expansion at sufficiently low energy densities in the presence of only pressureless matter.

We next show that under the above conditions \eqref{eqn:fcon}, the matter energy density $\rho_{\rm m}$ cannot decrease to zero as the Universe expands, but it can decrease to a nonzero minimum value which in turn also determines the minimum value of the deceleration parameter that the Universe could achieve from Eq. \eqref{qdef}. We see from Eq. \eqref{eq:rhoprime} that $H$ would never become null in the case $-1<\eta<0$ and $\alpha'>0$, and we also see from Eq. \eqref{eqn:conh} that in an expanding Universe, i.e., $H>0$, the matter energy density decreases with time, $\dot{\rho}_{\rm m}<0$, for $\rho_{\rm m}>\rho_{{\rm m},0} (-2\alpha'\eta)^{\frac{1}{1-2\eta}}$ and increases with time, $\dot{\rho}_{\rm m}>0$, for $\rho_{\rm m}<\rho_{{\rm m},0} (-2\alpha'\eta)^{\frac{1}{1-2\eta}}$. It follows that
\begin{equation}
\label{eqn:rmin1}
\rho_{\rm m,min}=\rho_{{\rm m},0}\left(\frac{1}{-2\alpha'\eta}\right)^{\frac{1}{2\eta-1}} \quad\textnormal{for}\quad -1<\eta<0.
\end{equation}
Substituting this into \eqref{qdef}, we find the minimum value of the deceleration parameter as
\begin{equation}
q_{\rm min}=\frac{2\eta^2-2\eta-1}{(2\eta-1)^2}\quad\textnormal{for}\quad -1<\eta<0.
\end{equation}
Next, we see from Eq. \eqref{eq:rhoprime} that $H$ would become null for $0<\eta<\frac{1}{2}$ and $\alpha'>0$ if $\rho_{m}$ could be null. However, we see from Eqs. \eqref{eqn:conh} and \eqref{eqn:hdot} that as the Universe expands ($H>0$), in the case $0<\eta<\frac{1}{2}$ and $\alpha'>0$ the matter energy density decreases with time as long as $\rho_{\rm m}<\rho_{{\rm m},0} \left(\frac{1-2\eta}{2\alpha'\eta}\right)^{\frac{1}{2\eta-1}}$, and eventually $\dot{\rho}_{\rm m}\rightarrow 0$ and $\dot{H}\rightarrow 0$ as $\rho_{\rm m}\rightarrow \rho_{{\rm m},0} \left(\frac{1-2\eta}{2\alpha'\eta}\right)^{\frac{1}{2\eta-1}}$. These signal that the matter energy density asymptotically approaches its minimum value
\begin{equation}
\label{eqn:rmin2}
\rho_{\rm m,min}=\rho_{{\rm m},0} \left(\frac{1-2\eta}{2\alpha'\eta}\right)^{\frac{1}{2\eta-1}}\quad\textnormal{for}\quad 0<\eta<\frac{1}{2},
\end{equation}
which gives [when substituted into Eq. \eqref{qdef}] the minimum value of the deceleration parameter that the Universe asymptotically approaches, as
\begin{equation}
\label{eqn:asym}
q_{\rm min}=-1 \quad\textnormal{for}\quad 0<\eta<\frac{1}{2}.
\end{equation}
According to this, although the EMPG modification would effectively behave like a phantom field $w'<-1$ for $0<\eta<\frac{1}{2}$, it would not lead to a big rip since the matter energy density does not diverge during the evolution of the Universe but asymptotically approaches a positive constant, which implies that the Hubble parameter also asymptotically approaches a positive constant, namely, the Universe would asymptotically approach exponential expansion with $q_{\rm min}=-1$ as given in the above equation. Finally, the fact that $a\rightarrow\infty$ as $\rho_{\rm m}\rightarrow \rho_{{\rm m},0} \left(\frac{1-2\eta}{2\alpha'\eta}\right)^{\frac{1}{2\eta-1}}$ in Eq. \eqref{eq:rh} [which is the solution we give below for $a(\rho_{\rm m})$] confirms our conclusions on the asymptotic behavior of our model for the case $0<\eta<\frac{1}{2}$.

In this paper, we do not intend to investigate the all possible solutions of the model, but rather those that satisfy the conditions we have discussed in this section for obtaining a viable cosmology. One may note that each different value of the parameter $\eta$ $-$which determines the power of the self-contraction of the energy-momentum tensor $-$would correspond to different theories of gravity, and hence a full mathematical analysis of the model for arbitrary values of $\eta$ or even for some particular values (such as that done, e.g., in Ref. \cite{Roshan:2016mbt} that corresponds to the case $\eta=1$ of our model in the presence of $\Lambda$) is beyond the scope of this paper. Here, our main purpose is to study whether viable cosmologies can be obtained from our model for certain ranges of the parameters $\alpha'$ (as well as $\alpha$) and $\eta$ estimated from observational data. Therefore, we continue here with the following solution for $a(\rho_{\rm m})$, satisfying the conditions given in Eqs. \eqref{eqn:fcon} and \eqref{cond1} and valid for $\rho_{\rm m}>\rho_{\rm m,min}$ given in Eqs. \eqref{eqn:rmin1} and \eqref{eqn:rmin2}:
\begin{align}
\frac{\rho_{\rm m}}{\rho_{{\rm m},0}}\left[\frac{1+\frac{2\eta\alpha'}{2\eta-1}\left(\frac{\rho_{\rm m}}{\rho_{{\rm m},0}}\right)^{2\eta-1}}{1+\frac{2\eta\alpha'}{2\eta-1}}\right]^{\frac{2\eta-2}{2\eta-1}} =a^{-3}.
\label{eq:rh}
\end{align} 
It should be noted here that we are not able to give an explicit solution for $\rho_{\rm m}(a)$ since it is not possible to isolate $\rho_{\rm m}$ except for a couple of particular cases of $\eta$ in this equation, which stands as one of the difficulties in the investigation of the model for arbitrary values of $\eta$. However, it might be immediately seen from this equation that pressureless matter in our model does not evolve as $a^{-3}$ except for $\eta=0,1$, and that $a\rightarrow\infty$ as $\rho_{\rm m}\rightarrow \rho_{{\rm m},0} \left(\frac{1-2\eta}{2\alpha'\eta}\right)^{\frac{1}{2\eta-1}}$ for $0<\eta<\frac{1}{2}$, in line with our expectation that the model would asymptotically approach exponential expansion in this case, in accordance with our discussion regarding Eq. \eqref{eqn:asym}.

\section{OBSERVATIONAL CONSTRAINTS}
 \label{obsanalysis}
 
In the previous section, we discussed in detail under what conditions our model could give viable cosmologies; we showed that, in the presence of only a matter source, our model can not only mimic the evolution of the Universe precisely as in the $\Lambda$CDM model at $\eta=0$, but it can also lead to an evolution similar to $w$CDM-like cosmology for $\eta\sim0$. However, in both cases the underlying physics of the EMPG model is entirely different than what we have in the $\Lambda$CDM and $w$CDM models. In this section, we investigate the observational constraints on the parameters $\alpha'$ (as well as $\alpha$) and $\eta$ by writing an approximated function for $a(\rho_{\rm m})$ [given in Eq. \eqref{eq:rh}] for $a\lesssim1$ (or $z\gtrsim0$) that would allow us to isolate $\rho_{\rm m}$, which we can then substitute into $H(\rho_{\rm m})$ [given in Eq. \eqref{eq:rhoprime}] to obtain $H(z)$ for the purpose of observational data analysis. In order to investigate the observational constraints on the parameters of the EMPG model, first we need to determine the matter energy density $\rho_{\rm m}$ of the model explicitly in terms of the cosmological redshift $z=-1+\frac{1}{a}$. However, $\rho_{\rm m}$ cannot be isolated explicitly in terms of $z$ from Eq. \eqref{eq:rh} except for a limited number of particular values of $\eta$. On the other hand, we find that the following explicit expression for the matter energy density $\rho_{\rm m}$ in terms of $z$ is a very good approximation to Eq. \eqref{eq:rh} for $\rho_{\rm m}/\rho_{{\rm m},0}\gtrsim1$ and $\eta\sim0$ (see Appendix \ref{app} for details):
\begin{align}
\label{eq:rhoo}
\rho_{\rm m}={\rho_{{\rm m},0}} [\beta(1+z)^3+1-\beta],
\end{align} 
where $\beta=\left(1+\frac{2\eta\alpha'}{2\eta-1}\right)^{\frac{2\eta-2}{2\eta-1}}$.
Finally, substituting Eq. \eqref{eq:rhoo} and the Hubble constant $H_0^2=\frac{\kappa}{3}(1+\alpha'){\rho_{{\rm m},0}}$ into Eq. \eqref{eq:rhoprime}, the approximated modified Friedmann equation in terms of redshift $z$ is
\begin{equation}
\begin{aligned}
\frac{H^2}{H_0^2}=&\frac{1}{1+\alpha'}\left[\beta (1+z)^3+1 - \beta\right] \\
&+\frac{\alpha'}{1+\alpha'}\left[\beta(1+z)^3+1 - \beta\right]^{2\eta},
\label{oo}
\end{aligned}
\end{equation}
subject to the conditions $\rho_{\rm m}\gtrsim{\rho_{{\rm m},0}}>\rho_{\rm m,min}$ (or $z\gtrsim0$) and the ones given in Eqs. \eqref{cond1} and \eqref{eqn:fcon}.

Having determined the evolution of the Hubble parameter in terms of $z$ in Eq. \eqref{oo},  now we can constrain the EMPG model parameters $\eta$, $\alpha'$ and $H_0$ with the observational data. For this purpose, we use the compilation of 28 Hubble parameter measurements spanning the redshift range $0.07 \leq z \leq 2.3$, as displayed in Table \ref{tab:Hz}. The $28$ $H(z)$ data points were compiled in Ref. \cite{Farooq:2013hq} to determine constraints on the parameters of various dark energy models. In a very recent paper \cite{Chen:2016uno}, the compilation of the 28 $H(z)$ points was utilized to determine the Hubble constant $H_0$ in four different cosmological models, including the $\Lambda$CDM and $w$CDM models. Following the same methodology, here we 
 constrain the
parameters $(H_0,\alpha',\eta)$ of the EMPG model by minimizing 
\begin{equation}
\label{eq:chi2Hz} \chi_{H}^2 (H_0,\alpha',\eta) =
\sum_{i=1}^{28}\frac{[H^{\rm th} (z_i; H_0,\alpha',\eta)-H^{\rm
obs}(z_i)]^2}{\sigma^2_{{\rm H},i}}
\end{equation}
for $28$ measured $H^{\rm obs} (z_i)$'s with variance $\sigma^2_{{\rm H},i}$
at redshift $z_i$ (as displayed in Table \ref{tab:Hz}), whereas $H^{\rm th}$ is the predicted value of $H(z)$ in the EMPG model. The parameter space $(H_0,\alpha',\eta)$ of the model is explored by using the Markov chain Monte Carlo method coded in the publicly available package COSMOMC \cite{Lewis:2002ah}. 
 \begin{table}[ht!]
 \captionsetup{justification=justified,singlelinecheck=false,font=footnotesize}
\begin{center}
\caption{Hubble parameter versus redshift data.}
\label{tab:Hz}
\begin{tabular}{cccc}
\hline\hline
$z$ & $H(z)$ &$\sigma_{H}$  &Reference\tnote{a}\\
    & (km s$^{-1}$ Mpc $^{-1}$) & (km s$^{-1}$ Mpc $^{-1}$)& \\
\tableline\\[-4pt]
0.070&    69&    19.6& \cite{Zhangetal2014}\\
0.090&    69&    12&    \cite{Simon:2004tf}\\
0.120&    68.6&    26.2&    \cite{Zhangetal2014}\\
0.170&    83&    8&    \cite{Simon:2004tf}\\
0.179&    75&    4&    \cite{Morescoetal2012}\\
0.199&    75&    5&    \cite{Morescoetal2012}\\
0.200&    72.9&    29.6&    \cite{Zhangetal2014}\\
0.270&    77&    14&    \cite{Simon:2004tf}\\
0.280&    88.8&    36.6&    \cite{Zhangetal2014}\\
0.350&    76.3&    5.6&    \cite{ChuangWang2013}\\
0.352&    83&    14&    \cite{Morescoetal2012}\\
0.400&    95&    17&    \cite{Simon:2004tf}\\
0.440&    82.6&    7.8&    \cite{Blakeetal2012}\\
0.480&    97&    62&    \cite{Sternetal2010}\\
0.593&    104&    13&    \cite{Morescoetal2012}\\
0.600&    87.9&    6.1&    \cite{Blakeetal2012}\\
0.680&    92&    8&    \cite{Morescoetal2012}\\
0.730&    97.3&    7.0&    \cite{Blakeetal2012}\\
0.781&    105&12&    \cite{Morescoetal2012}\\
0.875&    125&    17&    \cite{Morescoetal2012}\\
0.880&    90&40&    \cite{Sternetal2010}\\
0.900&    117&    23&    \cite{Simon:2004tf}\\
1.037&    154&    20&    \cite{Morescoetal2012}\\
1.300&168&    17&    \cite{Simon:2004tf}\\
1.430&    177&    18&    \cite{Simon:2004tf}\\
1.530&    140&    14&    \cite{Simon:2004tf}\\
1.750&    202&    40&    \cite{Simon:2004tf}\\
2.300&    224&    8&    \cite{Buscaetal2013}\\
\hline\hline
\end{tabular}
%\begin{tablenotes}
%\item[a]{Reference numbers:1. Simon et al. \cite{Simon:2004tf}, 2. Stern et al. \cite{Sternetal2010},
%3. Moresco et al. \cite{Morescoetal2012}, 4. Busca et al. \cite{Buscaetal2013},
%5. Zhang et al. \cite{Zhangetal2014}, 6. Blake et al. \cite{Blakeetal2012},
%7. Chuang \& Wang \cite{ChuangWang2013}}
%\end{tablenotes}
\end{center}
\end{table}

\begin{figure}[ht!]
  \captionsetup{justification=raggedright,singlelinecheck=false,font=footnotesize}
\includegraphics[width=0.48\textwidth]{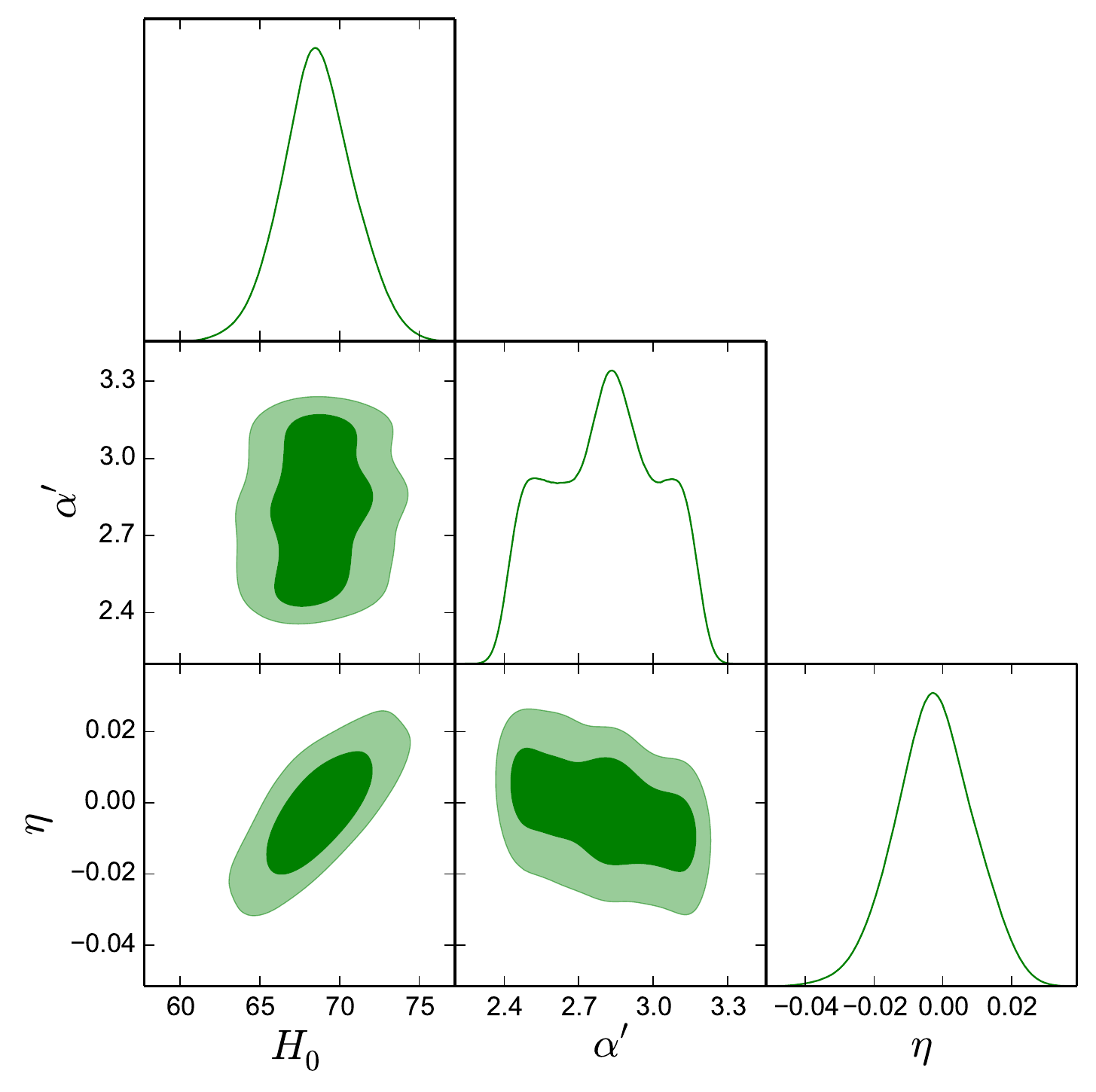}
\caption{1$\sigma$ and 2$\sigma$ confidence contours of the EMPG model parameters. Marginalized probability distributions of the individual parameters are also displayed.}
  \label{contour}
\end{figure}

Our results are presented in Table \ref{tab:results} and Fig. \ref{contour}.
\begin{table}[ht]
\captionsetup{justification=raggedright,singlelinecheck=false,font=footnotesize}
\begin{center}
\caption{Mean values of EMPG model parameters are displayed with 68\% and 95\% confidence levels (C.L.). We find $\chi^2_{\rm min}=17$.} 
\label{tab:results}
\begin{tabular}{cccc}
\hline\hline
Parameter & Mean & 68\% C.L. & 95\% C.L.  \\
\hline 
$H_0$    & $68.7$ & $[66.5, 70.9]$ & $[64.2, 73.2]$  \\[6pt]
$\alpha'$ & $2.80$ & $[2.54,3.05]$ & $[2.42,3.17]$  \\[6pt]
$\eta$    &$-0.003$&$ [-0.014,0.009]$ & $[-0.026,0.020]$ \\[6pt]
 \hline\\
\end{tabular}
\end{center}
\end{table}
We find that the values of the EMPG model parameters are $H_0=68.7^{+2.2+4.5}_{-2.2-4.5}$ km s$^{-1}$ Mpc $^{-1}$, $\alpha'=2.80^{+0.25+0.37}_{-0.26-0.38}$ and $\eta=-0.003^{+0.012+0.023}_{-0.011-0.023}$, where the error limits are at $1\sigma$ and $2\sigma$ confidence levels with $\chi^2_{\rm min}=17$. We first note that the values of $\alpha'$ and $\eta$ satisfy all of the conditions \eqref{cond1}, \eqref{cond2}, and \eqref{eqn:fcon} as described in the previous section for a viable cosmology. We note that the mean value of $\eta$ is almost equal to zero, signaling that our model$-$in light of observational data$-$predicts a $\Lambda$CDM-type background evolution at least up to the present time. The constraints on the$\Lambda$CDM model parameters read as $H_0=68.3^{+2.7+5.2}_{-2.6-5.1}$ km s$^{-1}$ Mpc $^{-1}$ and $\Omega_{\rm m,0}=0.276^{+0.032+0.072}_{-0.039-0.068}$, with error limits at the $1\sigma$ and $2\sigma$ confidence levels and $\chi^2_{\rm min}=17$ (see Table I of Ref. \cite{Chen:2016uno}). Recalling that the case $\eta=0$ corresponds to $\Lambda$CDM-type expansion with the transformation $\alpha'=\rho_{\Lambda}/\rho_{{\rm m},0}$ in our model, it is easy to deduce $H_0=68.3^{+2.7+5.2}_{-2.6-5.1}$ km s$^{-1}$ Mpc $^{-1}$, $\alpha'=2.623^{+0.596+1.185}_{-0.376-0.750}$ for $\eta=0$ (fixed). It is noteworthy that the errors of $\alpha'$ become larger when $\eta$ is fixed to zero, which may be interpreted as the data suggesting that fixing $\eta$ to zero is not preferable. On the other hand, we see that the observational data of the Hubble parameter measurements fit equally well to the approximated EMPG model, in contrast with the $\Lambda$CDM model, and predict a similar value of $H_0$ in both models.

\begin{figure}[t!]
\captionsetup{justification=raggedright,singlelinecheck=false,font=footnotesize}
\includegraphics[width=0.5\textwidth]{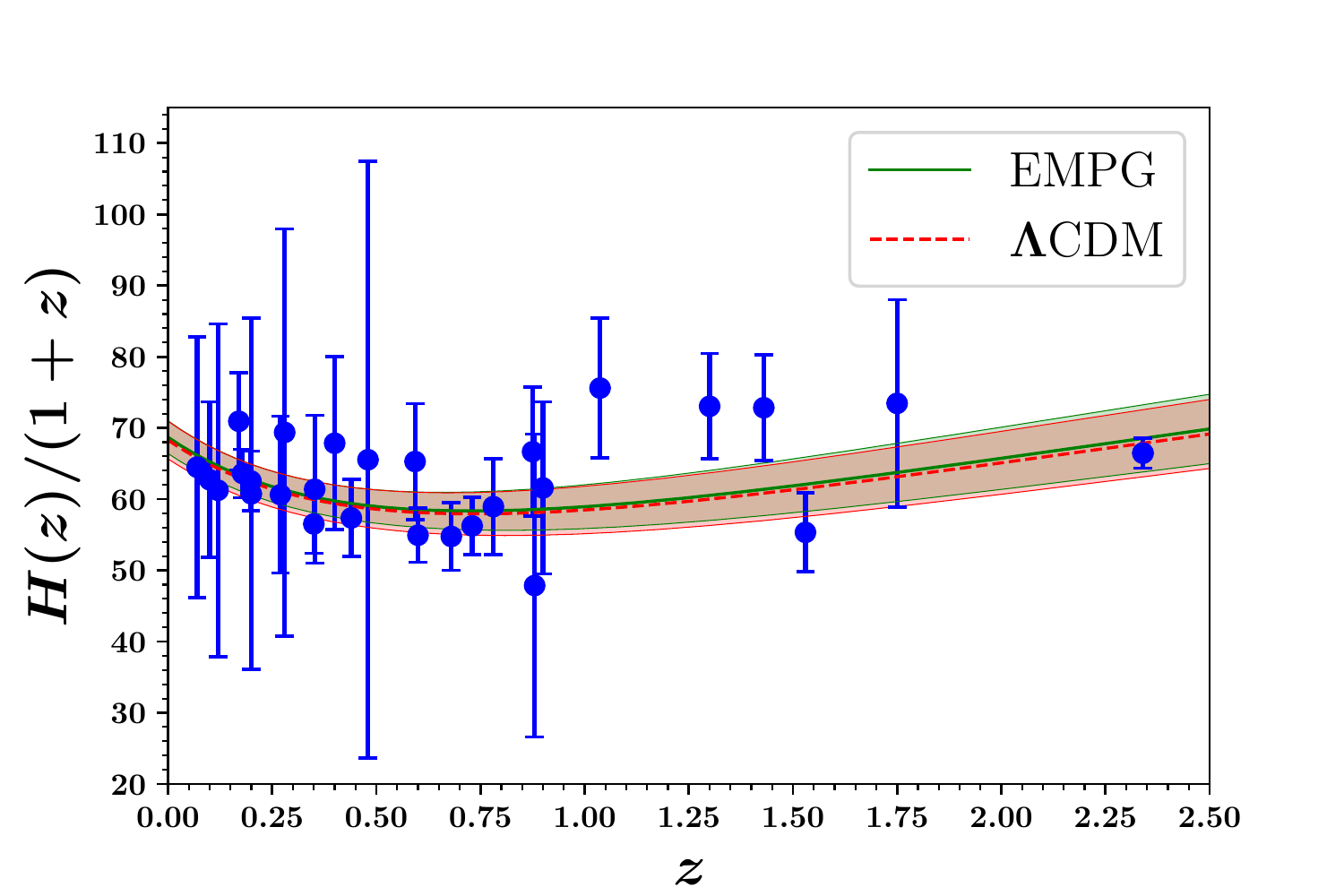}
  \caption{$H(z)/(1+z)$ curves of the EMPG and $\Lambda$CDM models are plotted with the mean values and 1$\sigma$ error regions of the model parameters. The $28$ $H(z)$ data points of Table 1 are also displayed.}
  \label{Hzfit}
\end{figure}

\begin{figure}[!bt]
\captionsetup{justification=raggedright,singlelinecheck=false,font=footnotesize}
\includegraphics[width=0.5\textwidth]{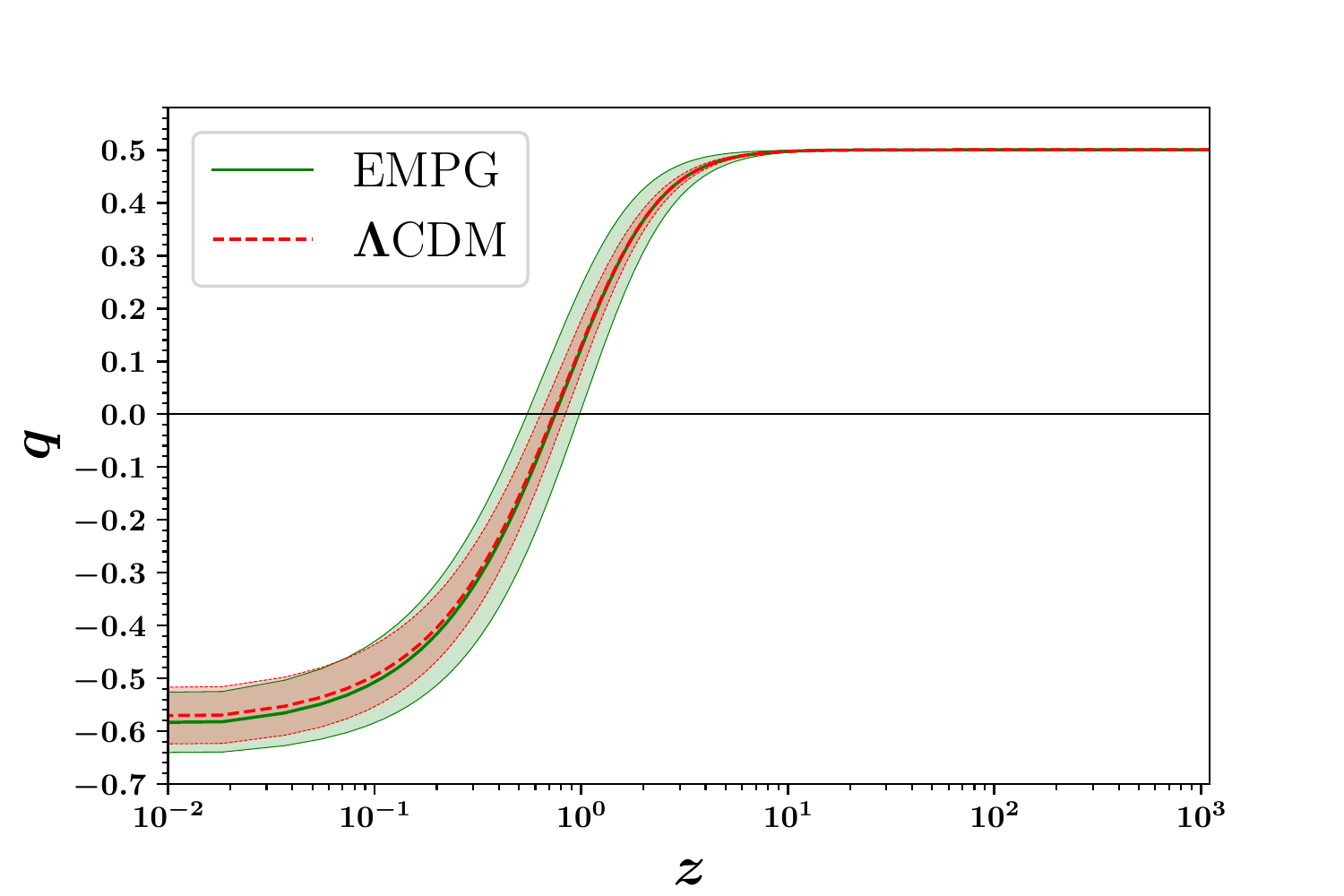}
  \caption{The mean value deceleration parameter curves with 1$\sigma$ error regions of the EMPG and $\Lambda$CDM models are plotted in the redshift range $(0,1100)$ (log scale on the $z$-axis) with the mean values of the model parameters.  }
  \label{qz}
\end{figure}

\begin{figure}[!bt]
\captionsetup{justification=raggedright,singlelinecheck=false,font=footnotesize}
\includegraphics[width=0.5\textwidth]{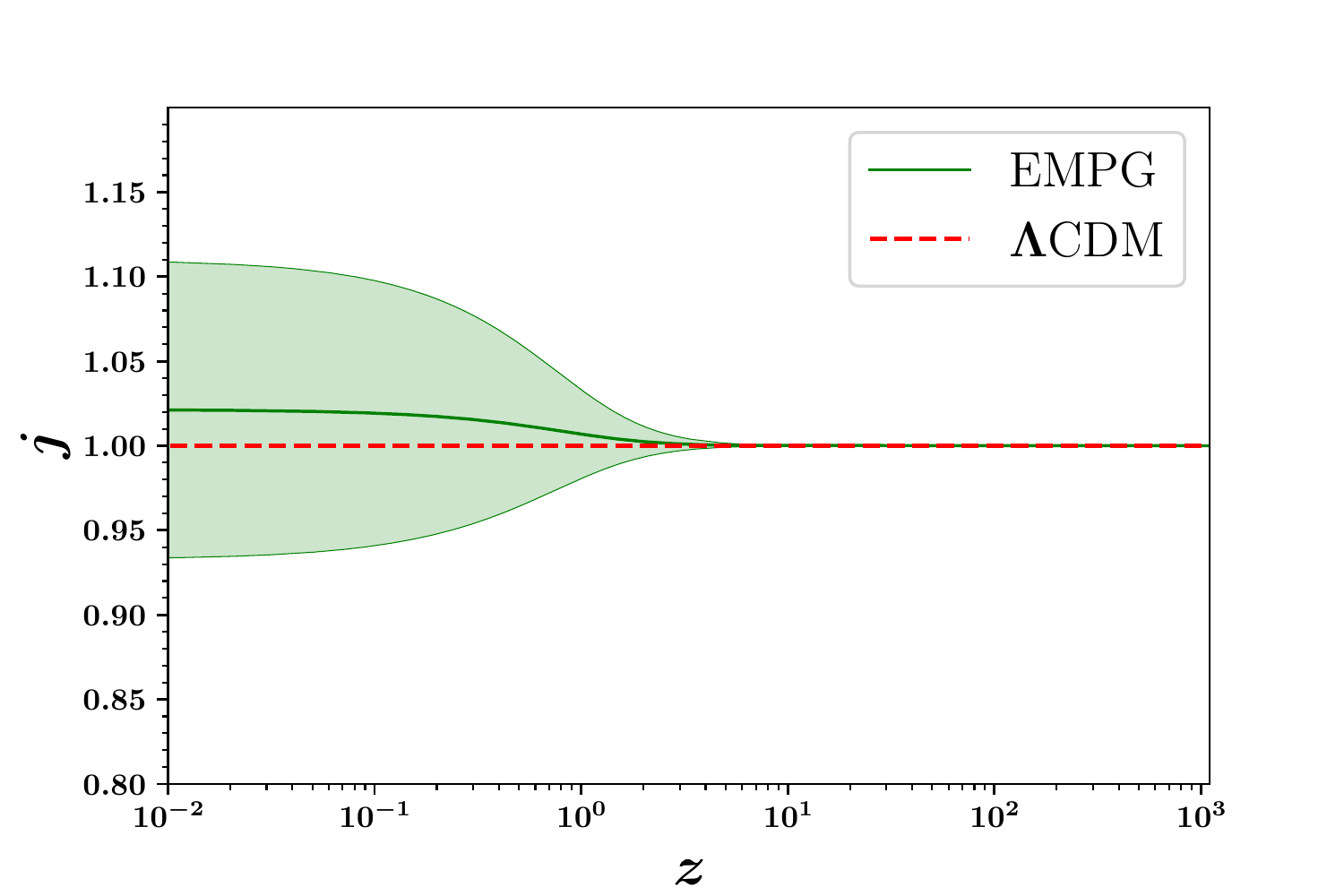}
  \caption{The mean value jerk parameter curves with 1$\sigma$ error regions of the EMPG and $\Lambda$CDM models are plotted in the redshift range $(0,1100)$ (log scale on  $z$-axis) with the mean values of the model parameters.}
  \label{jz}
\end{figure}

In Fig. \ref{Hzfit}, we show the mean value $H(z)/(1+z)$ curves with the 1$\sigma$ error region of the EMPG and $\Lambda$CDM  models in the redshift range $0\leq z\leq 2.5$ along with the data points from Table \ref{tab:Hz}. We see that the $H(z)/(1+z)$ curves of the two models overlap, indicating a similar evolution of the Universe in the two models.
The mean value evolution trajectories of the deceleration parameter $q=-1+(1+z)\frac{H'}{H}$ and jerk parameter $j=1-2(1+z)\frac{H'}{H}+(1+z)^2\left(\frac{H'^2}{H^2}+\frac{H''}{H}\right)$ with 1$\sigma$ error regions of the EMPG model and $\Lambda$CDM model (yielding a constant jerk parameter equal to unity, $j_{\Lambda{\rm CDM}}=1$) in the redshift range $0\leq z\leq 1100$ (log scale on the $z$ axis), are shown in Figs. \ref{qz} and \ref{jz}, respectively.
Again, we observe a similar evolution of the Universe in the two models in the whole redshift range $0\leq z\leq 1100$. Thus, the EMPG and $\Lambda$CDM models provide similar descriptions of the dynamics and kinematics of the Universe up to the present time, in light of observational data from Hubble parameter measurements, despite the fact that the underlying physics of the EMPG model is completely different from that of the $\Lambda$CDM model. Thus, the observed Hubble data supports the new physics employed in this work to develop the EMPG model, which would deviate in various ways from the $\Lambda$CDM model depending on the sign of the $\eta$ (as the data allow both negative and positive values of $\eta$). However, we do not discuss the future dynamics of the Universe in the EMPG model in this paper.

\begin{figure}[h]
\captionsetup{justification=raggedright,singlelinecheck=false,font=footnotesize}
\includegraphics[width=0.49\textwidth]{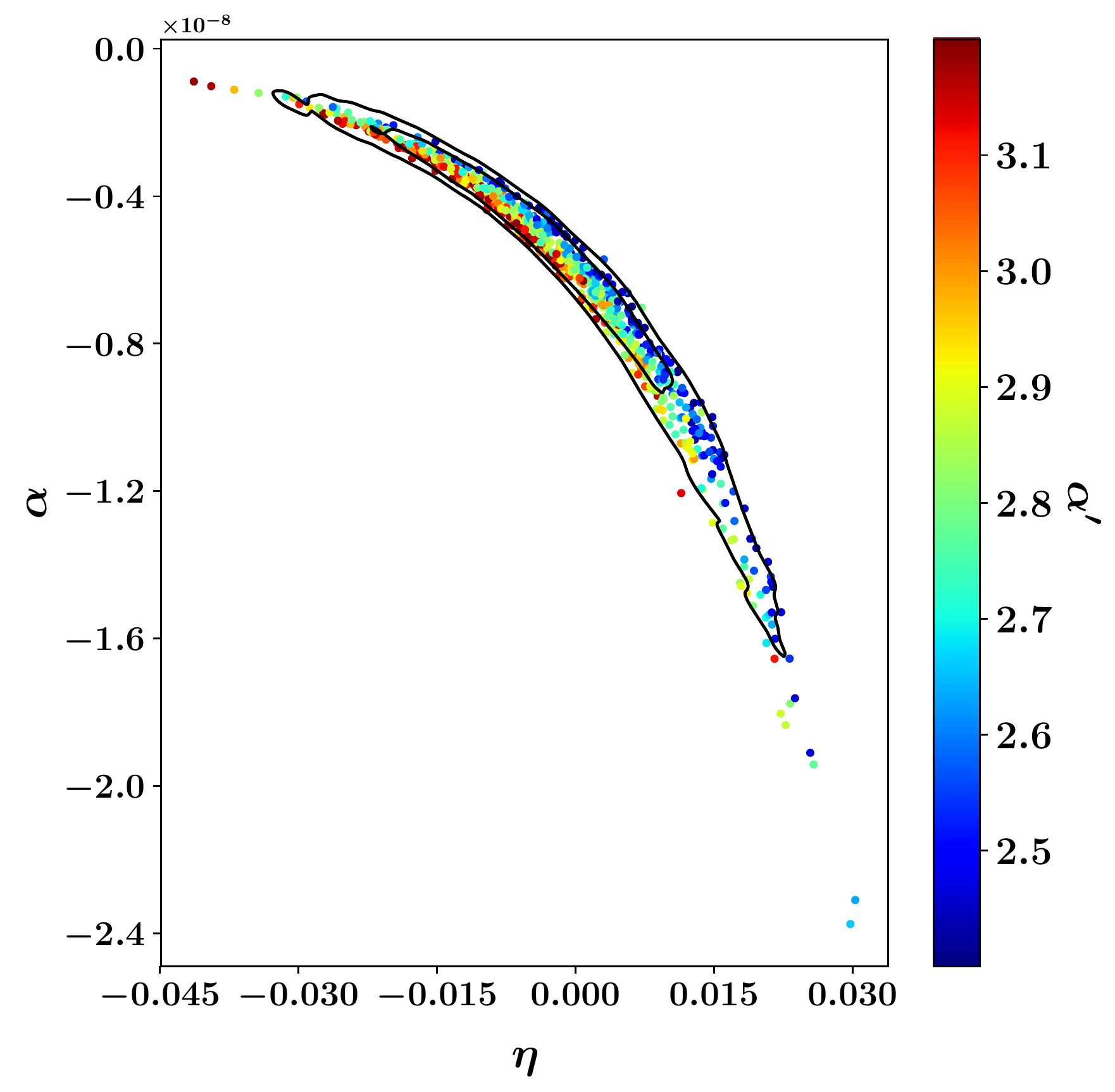}
  \caption{Constraints in the $\eta$-$\alpha$ plane with 1$\sigma$ and 2$\sigma$ confidence level contours from $H(z)$ data in the EMPG model where the samples are colored by the parameter $\alpha'$.}
  \label{alpahaeta}
\end{figure}

Finally, to find the constraints on the parameter $\alpha$$-$which like $\eta$, is a true constant of the model that appears in the action of the EMPG model$-$ we first obtain the equation
\begin{equation}
\label{eqn:Aprime}
\alpha=\frac{\alpha'}{2\eta-1}\left[\frac{3H_0^2}{\kappa(\alpha'+1)}\right]^{1-2\eta}
\end{equation}
in the presence of only a matter source upon using Eq. \eqref{eq:presprime} with Eq. \eqref{eq:alphaprime} by substituting $w=0$. Hence, using the constraints on the parameters given in Table \ref{tab:results} in Eq. \eqref{eqn:Aprime} we obtain the constraint on the parameter $\alpha$ as
\begin{equation}
\alpha = - 0.60^{+0.50}_{-0.69}\times10^{-8} (\rm erg/cm^3)^{1-2\eta}\quad (95\%\, \rm C.L.).
\label{calpha}
\end{equation}
One may note that the units of $\alpha$ depend on $\eta$, which indicates that each different value of $\eta$ should be considered as another gravity theory. In Fig. \ref{alpahaeta} we present the three dimensional constraints in the $\eta$-$\alpha$ plane from $H(z)$ data, where the samples are colored by the parameter $\alpha$.
It may be noted that we did not take the parameter $\alpha$ into account in the cosmological model that we  studied in the presence of only pressureless matter, but rather the more useful parameter $\alpha'$ [defined in Eq. \eqref{eq:alphaprime}], particularly for a monofluid cosmological solution of the model. However, the true constants of the EMPG model are the parameters $\alpha$ and $\eta$ that appear in the action \label{eq:action} of the EMPG model. In other words, the constraints we give on $H_0$, $\eta$, and $\alpha'$ above constitute the constraints on the cosmological model in the presence of only a pressureless source, while the constraints on $\eta$ and $\alpha$ constitute the constraints on the parameters of the EMPG model by means of the cosmological model under consideration. One may try to obtain constraints on $\eta$ and $\alpha$ from noncosmological physics, e.g., neutron stars, parametrized post-Newtonian parameters etc. In this case it might not be possible to properly define a parameter $\alpha'$ corresponding to the one we defined in this paper. On the other hand, one may try to obtain constraints on $\eta$ and $\alpha$ from, for instance, big bang nucleosynthesis, during which the Universe would be considered to be filled with only radiation ($w=1/3$). In this case one can properly use the $\alpha'$ given in Eq. \eqref{eq:alphaprime} by substituting $w=1/3$. However, even in this case, one cannot compare the constraint that would be found for $\alpha'$ for radiation with the one we give here for pressureless fluid in Table \ref{tab:results}; rather, one should deduce the constraint on $\alpha$ and then compare that with the one we give in Eq. \eqref{calpha}. A good demonstration of this point can be given by the following example. We know from Eq. \eqref{eq:alphaprime} that it is not necessary for $\alpha'$ to be the same for different fluids since it depends on the EoS parameter $w$ of the considered fluid. We found here that $\alpha'>0$ for pressureless matter ($w=0$) and then deduced that $\alpha<0$. However, if we use the above constraints we obtained for $\eta$ and $\alpha$, we see from \eqref{eq:alphaprime} that $\alpha'<0$ for radiation ($w=1/3$). Hence, the constraints we obtained on $\eta$ and $\alpha$ using data obtained from the late Universe predict that the EMPG modification would lower the value of the Hubble parameter in the radiation-dominated Universe$-$for instance, when BBN took place$-$though the EMPG modification would be completely negligible at the energy scales of BBN.

\section{Concluding remarks and future perspectives}

In this study, we have proposed a modified theory of gravitation constructed by the addition of the term $f(T_{\mu\nu}T^{\mu\nu})$ to the EH action of GR, and elaborated a particular case $f(T_{\mu\nu}T^{\mu\nu})=\alpha(T_{\mu\nu}T^{\mu\nu})^{\eta}$ (where $\alpha$ and $\eta$ are real constants dubbed EMPG. We have further investigated the EMPG model on theoretical and observational grounds by considering the cosmological evolution that it predicts. We have discussed the conditions under which it leads to viable cosmologies, and showed that there are ranges of the parameters of the EMPG model in which it approaches standard GR at high energy densities and the accelerated expansion at sufficiently low energy densities (namely, in the present Universe) without invoking a cosmological constant or any other dark energy source. We have shown that, in the presence of only a matter source, EMPG can give rise to not only precisely the same background evolution as the $\Lambda$CDM model when $\eta=0$, but also an evolution similar to that of the $w$CDM model when $\eta\sim0$. However, in both cases the underlying physics of the EMPG-driven cosmologies are entirely different than what we have in the $\Lambda$CDM and $w$CDM models. For instance, we introduced neither $\Lambda$ nor a dark energy source to drive the late-time acceleration of the Universe. Our model alleviates the cosmological constant problem that arises when a cosmological constant is introduced as in the $\Lambda$CDM model, and the issue of introducing an \textit{ad hoc} scalar field that can lead to quintessence and/or phantom (leading additional severe problems) dark energy source with a constant or slightly varying equation-of-state parameter as in the $w$CDM-like models. The constraints on the EMPG model parameters from the recent compilation of $28$ Hubble parameter measurements reveal that the EMPG model describes an evolution of the Universe similar to that in the $\Lambda$CDM model.

We have shown that our modification of the EH action that leads to the EMPG action becomes negligible at sufficiently high energy densities leaving the standard cosmology based on GR untouched for times earlier than the time of onset of the late-time acceleration of the Universe. This means that if we had added the term $f(T_{\mu\nu}T^{\mu\nu})=\alpha(T_{\mu\nu}T^{\mu\nu})^{\eta}$  to the Starobinsky action \cite{Starobinsky:1980te} rather than the EH action, then this term would be even more negligible at inflationary energy scales and hence the Starobinsky inflation model \cite{Starobinsky:1980te} (which is the leading inflationary model supported by the most recent cosmological observations \cite{Ade:2015lrj}) would also be left untouched. Hence, a modified gravity model such as the Starobinsky action$+\alpha(T_{\mu\nu}T^{\mu\nu})^{\eta}$ with $\eta\sim0$ and $\alpha\sim2.5$ would be able to successfully describe the complete history of the Universe  (see Appendix \ref{MGR} for more details).

Our findings in this paper are promising enough to justify further study of EMPG, as it is apparently a very rich topic. In our detailed discussions here we have mainly considered a pressureless matter source and its evolution throughout the history of Universe, and discussed the late-time acceleration of the Universe. Hence, one direction may be to extend our analysis by considering effects of other known cosmological fluids on the evolution of the Universe. For instance, radiation could be studied in order to obtain constraints on the parameters of the EMPG from BBN processes that took place when the Universe was radiation dominated. Of course, the inclusion of a positive $\Lambda$ in our model is also of further interest to us since in this case the model would provide much richer dynamics (e.g. Ref. \cite{Board:2017ign}). In particular, if we consider the case $\eta>\frac{1}{2}$, then EMPG would alter the dynamics of the early Universe, but lead to an evolution of the Universe like that in the $\Lambda$CDM model after a certain amount of time (see Refs. \cite{{Roshan:2016mbt},ACEK} for EMSG, i.e., ${\rm EMPG}_{\eta=1}$). In this case, one may need to study not only the cosmological dynamics but also the physical processes/astrophysical objects relevant to high energy densities such as BBN/neutron stars, which in turn can be used jointly to constrain the parameters of the model without suffering from degeneracy issues. Another direction may be to consider functions other than $f(T_{\mu\nu}T^{\mu\nu})=\alpha(T_{\mu\nu}T^{\mu\nu})^{\eta}$ that lead to the EMPG we studied here; for instance, $f(T_{\mu\nu}T^{\mu\nu})=\alpha\log(T_{\mu\nu}T^{\mu\nu})$ would give $\log(\rho^2)$-type modifications to Einstein's field equations, as in  Ref. \cite{Chavanis:2016pcp}. The essence of constructing modified gravity models is to redefine the coupling of the energy-momentum tensor to the spacetime geometry. However, this is usually done by modifying the action that leads to modifications on the left-hand side of the Einstein's field equations, in contrast to what we have done here. Manipulating the introduction of the energy-momentum tensor into the action would lead to modified Einstein's field equations that might not be possible to obtain or to easily find from the Einstein's field equations with a modification on the left-hand side while the energy density and pressure of a fluid appear on the right-hand side of the equations as usual. Hence, it would be interesting to investigate the $f(R)$ correspondence of $T_{\mu\nu}T^{\mu\nu}$-type modifications in the metric and Palatini formalisms, similarly to, for instance, how the Ricci-tensor-powered gravity was studied in the Palatini variational approach regarding the powered form of the Ricci tensor contraction in Ref. \cite{Li:2007xw}. Finally, the geometrical counterpart of  $T_{\mu\nu}T^{\mu\nu}$-type modifications can also be studied, such as the powered $G_{\mu\nu}G^{\mu\nu}$ term, and even the geometry and source can be mixed by considering terms like $G_{\mu\nu}T^{\mu\nu}$. For instance, if the scalar field EMT is considered as the source, the model should correspond to a very popular subclass of Horndeski scalar-tensor theories \cite{Horndeski:1974wa}, where the scalar field is nonminimally  coupled with gravity through the Einstein tensor such that $G_{\mu\nu}\partial^{\mu}\phi\partial^{\nu}\phi$.
\begin{acknowledgements}
We thank to R.C. Nunes for constructive discussions. \"{O}.A. acknowledges the support by the Science Academy in scheme of the Distinguished Young Scientist Award  (BAGEP).  N.K. acknowledges the post-doctoral research support she is receiving from the {\.I}stanbul Technical University. S.K. gratefully acknowledges the support from SERB-DST project No. EMR/2016/000258.
\end{acknowledgements}
\appendix
\section{Approximation procedure}
\label{app}
\begin{figure}[!htb]
  \captionsetup{justification=raggedright,singlelinecheck=false,font=footnotesize}
\includegraphics[width=0.49\textwidth]{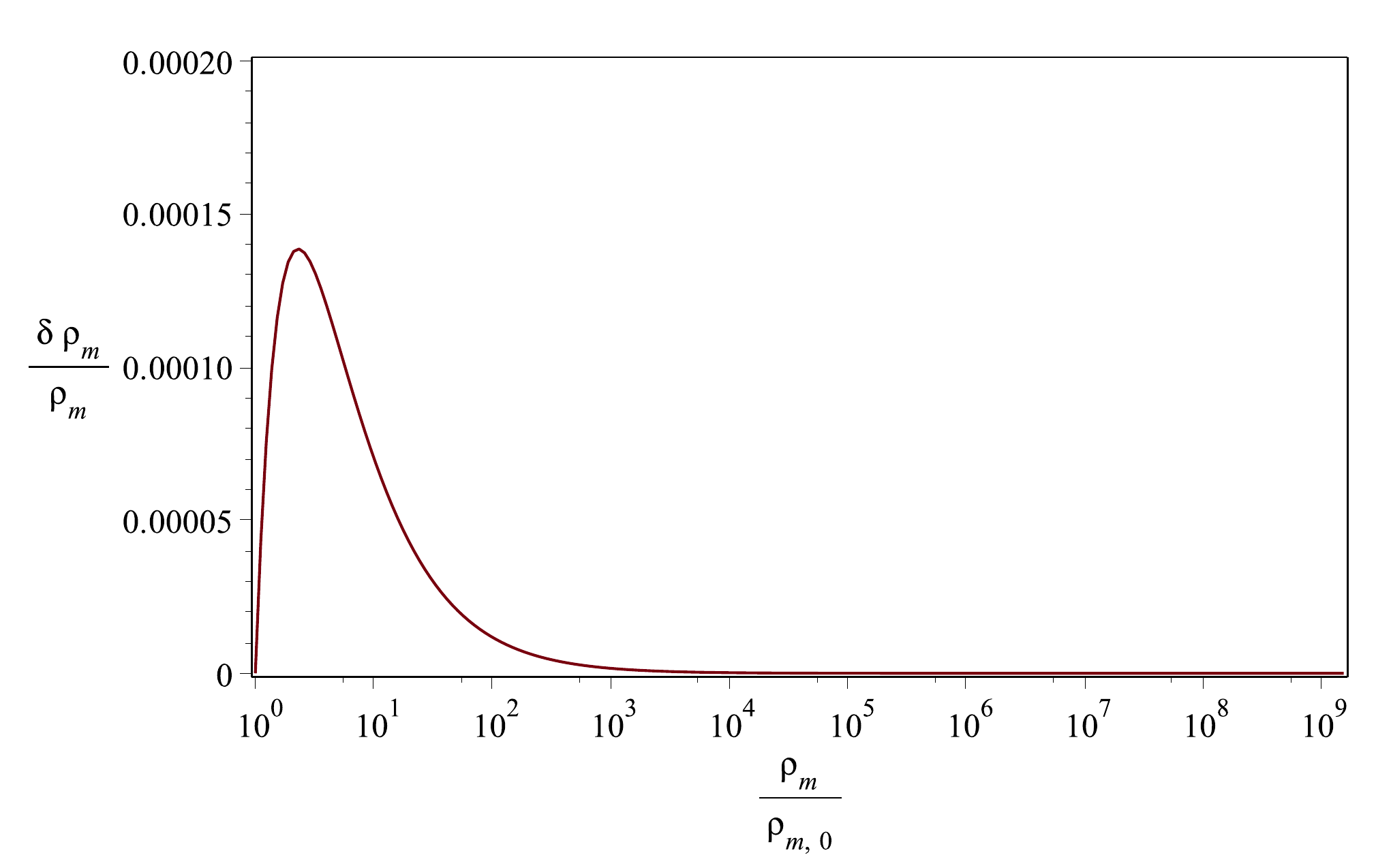}
  \caption{The relative error $\delta\rho_{\rm m}/\rho_{\rm m}$ versus $\rho_{\rm m}/\rho_{{\rm m},0}$ from $\rho_{\rm m}/\rho_{{\rm m},0}=1$ $(z=0)$ to $\rho_{\rm m}/\rho_{{\rm m},0}=10^9$ $(z\sim1100)$ for the mean values $\alpha'=2.8$, $\eta=-0.003$ given in Table \ref{tab:results}.}
  \label{rhofit}
\end{figure}

The matter energy density $\rho_{\rm m}$ appears implicitly in Eq. \eqref{eq:rh},
\begin{align}
\label{eqn:apx1}
\frac{\rho_{\rm m}}{\rho_{{\rm m},0}}\left[\frac{1+\frac{2\eta\alpha'}{2\eta-1}\left(\frac{\rho_{\rm m}}{\rho_{{\rm m},0}}\right)^{2\eta-1}}{1+\frac{2\eta\alpha'}{2\eta-1}}\right]^{\frac{2\eta-2}{2\eta-1}} =(1+z)^{3},
\end{align}
 and it cannot be isolated in terms of redshift $z$. Therefore, it is not possible to explicitly write the exact function for the Hubble parameter $H(z)$  by substituting $\rho_{\rm m}(z)$ in Eq. \eqref{eq:rhoprime}, but a good approximation can be found. To do so, we first note that Eq. \eqref{eqn:apx1} can be written as
\begin{align}
\label{eqn:fs}
\left({1+\frac{2\alpha'\eta}{2\eta-1}}\right)^{-\frac{2\eta-2}{2\eta-1}}\rho_{\rm m}\approx \rho_{{\rm m},0}(1+z)^{3}\;\;\textnormal{for}\;\; \rho_{\rm m}\gg\rho_{{\rm m},0},
\end{align}
provided that $\eta<\frac{1}{2}$ in accordance with the conditions given in Eq. \eqref{eqn:fcon}. Of course, this approximation is not good in the vicinity of $z= 0$, and we need to improve it. To do so, we first note that the approximation error (the deviation from the true value) at $z=0$ is $\left({1+\frac{2\alpha'\eta}{2\eta-1}}\right)^{-\frac{2\eta-2}{2\eta-1}}\rho_{{\rm m},0}-\rho_{{\rm m},0}$. Notice that, in this case, we do not get $\rho_{\rm m}=\rho_{{\rm m},0}$ at $z=0$, but $\rho_{\rm m}=\left({1+\frac{2\alpha'\eta}{2\eta-1}} \right)^{-\frac{2\eta-2}{2\eta-1}}\rho_{{\rm m},0}$. Therefore, subtracting this approximation error from the left-hand side of Eq. \eqref{eqn:fs} would not only give the actual value of $\rho_{\rm m}$ when $z=0$, but also decrease the approximation error when $\rho_{\rm m}\sim\rho_{{\rm m},0}$, viz., when $z\sim0$. Accordingly, compensating the approximation error at $z=0$ in Eq. \eqref{eqn:fs}, we reach the following improved approximation:
\begin{equation}
\begin{aligned}
&\left({1+\frac{2\alpha'\eta}{2\eta-1}}\right)^{-\frac{2\eta-2}{2\eta-1}}\rho_{\rm m}\\
&-\left[\left({1+\frac{2\alpha'\eta}{2\eta-1}}\right)^{-\frac{2\eta-2}{2\eta-1}}\rho_{{\rm m},0}-\rho_{{\rm m},0}\right]\approx\rho_{{\rm m},0}(1+z)^3,
\end{aligned}
\end{equation}
which is an approximation both for $\rho_{\rm m}\gg\rho_{{\rm m},0}$ and $\rho_{\rm m}\sim\rho_{{\rm m},0}$. In addition, it allows us to isolate $\rho_{\rm m}$ in terms of $z$ as
\begin{equation}
\begin{aligned}
\label{eqn:apxr}
\rho_{\rm m}\approx\rho_{\rm m,approx}=&\rho_{{\rm m},0} \left[1-\left({1+\frac{2\alpha'\eta}{2\eta-1}}\right)^{\frac{2\eta-2}{2\eta-1}}\right]\\
&+\rho_{{\rm m},0} \left({1+\frac{2\alpha'\eta}{2\eta-1}}\right)^{\frac{2\eta-2}{2\eta-1}} (1+z)^3.
\end{aligned}
\end{equation}
In Fig. \ref{rhofit}, we depict the relative error between the true value of the matter energy density $\rho_{\rm m}$ from Eq. \eqref{eq:rh} [or Eq. \eqref{eqn:fs}] and the approximated value of the matter energy density $\rho_{\rm m,approx}$ from Eq. \eqref{eqn:apxr}$-$namely, $\frac{\delta\rho_{\rm m}}{\rho_{\rm m}}=\frac{\rho_{\rm m,approx}-\rho_{\rm m}}{\rho_{\rm m}}$ versus $\rho_{\rm m}/\rho_{{\rm m},0}$ from $\rho_{\rm m}/\rho_{{\rm m},0}=1\; (z=0)$ to $\rho_{\rm m}/\rho_{{\rm m},0}=10^9\; (z\sim1100)$ $-$while using the mean values $\alpha'=2.8$, $\eta=-0.003$ from Table \ref{tab:results}. We see that the relative error is indeed negligible, as the maximum relative error is only $0.01$ percent while it is much smaller for $\rho_{\rm m}\sim\rho_{{\rm m},0}$ and $\rho\gg\rho_{{\rm m},0}$.

\section{The complete history of the Universe via the Starobinsky action extension}
\label{MGR}
EMPG is studied on theoretical as well as observational grounds leading to the late-time acceleration of the Universe with only pressureless matter while keeping the successes of standard general relativity at earlier times; however, inflation is not considered. For completeness, we may consider the modification that gives EMPG in the Starobinsky gravity action \cite{Starobinsky:1980te} instead of that in the Einstein-Hilbert action \eqref{action} as follows:
\begin{align}
S=\int \left[\frac{1}{2\kappa}\left(R+\frac{R^2}{6M_{\rm p}^2}\right)+\alpha(T_{\mu\nu}T^{\mu\nu})^{\eta}+\mathcal{L}_{\rm m}\right]\sqrt{-g}d^4x,
\label{eq:staroaction}
\end{align}
The field equations become
\begin{equation}
\begin{aligned}
\left(1+\frac{R}{3M_{\rm p}^2}\right)G_{\mu\nu}+\left(g_{\mu\nu}\Box-\nabla_{\mu}\nabla_{\nu}\right)\left(1+\frac{R}{3M_{\rm p}^2}\right)
\\ 
=\kappa T_{\mu\nu} +\kappa \alpha (T_{\sigma\epsilon}T^{\sigma\epsilon})^{\eta}\left[g_{\mu\nu}-2\eta\frac{\theta_{\mu\nu}}{T_{\sigma\epsilon}T^{\sigma\epsilon}}\right].
\label{starofieldeq}
\end{aligned}
\end{equation}
Then, using Eqs. \eqref{thetafrw} and \eqref{trace} as well as the metric \eqref{RW} in the field equations \eqref{starofieldeq}, we get the following set of two independent differential equations:
\begin{equation}
\begin{aligned}
3\left(1+\frac{R}{3M_{\rm p}^2}\right)H^2+\frac{H\dot{R}}{M_{\rm p}^2}-\frac{R^2}{12M_{\rm p}^2}=\kappa\rho+\kappa'  \rho_0\left(\frac{\rho}{\rho_0}\right)^{2\eta} , 
\label{eq:starhoprime} 
\end{aligned}
\end{equation}
\begin{equation}
\begin{aligned}
\left(-2\dot{H}-3H^2\right)\left(1+\frac{R}{3M_{\rm p}^2}\right)-\frac{H\dot{R}+\ddot{R}}{3M_{\rm p}^2}+\frac{R^2}{12M_{\rm p}^2}
\\=\kappa w\rho+ \frac{\kappa' \rho_0}{2\eta-1+\frac{8w\eta}{3w^2+1}} \left(\frac{\rho}{\rho_0}\right)^{2\eta}.\\
\label{eq:stapresprime}
\end{aligned}
\end{equation}
We note that the terms that appear in the field equations given in Eqs. \eqref{eq:starhoprime} and \eqref{eq:stapresprime} due to the $R^2$ term in the action \eqref{eq:staroaction} become effective for large values of $H$, namely, in the very early Universe at very high energy densities where the terms that appear in the field equations due to the modification $\alpha(T_{\mu\nu}T^{\mu\nu})^{\eta}$ in the action, are completely negligible for $\eta<\frac{1}{2}$. It is easy to see in this case that, for $\eta\sim0$ and $\alpha \sim-10^{-8} (\rm erg/cm^3)^{1-2\eta}$  (or $\kappa'\sim2.5\kappa$ in the presence of only dust) as found in this study, at inflationary energy scales the terms that appear in the field equations due to $R^2$ and the usual matter Lagrangian $\mathcal{L}_{\rm m}$ in the action would be effective, while the terms due to $R$ and the EMP modification would be negligible, so that $R^2$ inflation$-$which is the preferred model in light of the latest observational data by the Planck Collaboration \cite{Ade:2015lrj}$-$would take place during the early Universe in our model. Then, at intermediate energy scales (at the time of BBN, recombination, etc.), the terms that appear in the field equations due to the terms $R$ and $\mathcal{L}_{\rm m}$ in the action would be effective, but the terms due to the terms $R^2$ and  $\alpha(T_{\mu\nu}T^{\mu\nu})^{\eta}$ would be negligible so that the standard cosmology would work. Finally, at sufficiently low energy density scales the terms that appear in the field equations due to the terms $R$ and $\alpha(T_{\mu\nu}T^{\mu\nu})^{\eta}$ in addition to the $\mathcal{L}_{\rm m}$ in the action would be effective, while the terms in the field equations due to $R^2$ in the action would be negligible so that the Universe would start to accelerate at later times. Thus, the unification of EMPG with Starobinsky gravity would provide us with a complete history of the Universe. For instance, the Starobinsky model within the $f(R,T)$ formalism was studied in Ref. \cite{Moraes:2016zdw}, where the radiation$-$, matter$-$ and dark energy-dominated eras as well as the transition between these phases were studied.


\begin{thebibliography}{99}
%%%%%%%%%%%%%%%%%%%%%%%%%%%%%%%%
%\cite{Starobinsky:1980te}
\bibitem{Starobinsky:1980te} 
  A.~A.~Starobinsky, A new type of isotropic cosmological models without singularity,
  Phys.\ Lett.\ B {\bf 91}, 99 (1980).

  %\cite{Guth:1980zm}
\bibitem{Guth:1980zm} 
  A.~H.~Guth, The inflationary universe: A possible solution to the horizon and flatness problems,
  Phys.\ Rev.\ D {\bf 23}, 347 (1981).
  %doi:10.1103/PhysRevD.23.347
  %%CITATION = doi:10.1103/PhysRevD.23.347;%%
  %6444 citations counted in INSPIRE as of 07 Sep 2017

%\cite{Linde:1981mu}
\bibitem{Linde:1981mu} 
  A.~D.~Linde, A new inflationary universe scenario: A possible solution of the horizon, flatness, homogeneity, isotropy and primordial monopole problems,
  Phys.\ Lett.\  B {\bf 108}, 389 (1982).
  %doi:10.1016/0370-2693(82)91219-9
  %%CITATION = doi:10.1016/0370-2693(82)91219-9;%%
  %3973 citations counted in INSPIRE as of 07 Sep 2017
  
  %\cite{Albrecht:1982wi}
\bibitem{Albrecht:1982wi} 
  A.~Albrecht and P.~J.~Steinhardt, Cosmology for grand unified theories with radiatively induced symmetry breaking, Phys.\ Rev.\ Lett.\  {\bf 48}, 1220 (1982).
  %doi:10.1103/PhysRevLett.48.1220
  %%CITATION = doi:10.1103/PhysRevLett.48.1220;%%
  %3528 citations counted in INSPIRE as of 07 Sep 2017
  
  %\cite{Baumann:2009ds}
   \bibitem{Baumann:2009ds} 
D.~Baumann, Inflation, in TASI 2009: Physics of the Large and the Small, edited by C. Csaki and S. Dodelson (World Scientific, Singapore, 2011), p. 523, [arXiv:0907.5424 [hep-th]].

  
  %\cite{Linde:2014nna}
\bibitem{Linde:2014nna} 
  A.~Linde, Inflationary cosmology after Planck 2013,
  %doi:10.1093/acprof:oso/9780198728856.003.0006
  arXiv:1402.0526.
  %%CITATION = doi:10.1093/acprof:oso/9780198728856.003.0006;%%
  %186 citations counted in INSPIRE as of 07 Sep 2017
  
        %\cite{Ade:2015lrj}
\bibitem{Ade:2015lrj} 
  P.~A.~R.~Ade {\it et al.} [Planck Collaboration], Planck 2015 results. XX. Constraints on inflation,
  Astron.\ Astrophys.\  {\bf 594}, A20 (2016), [arXiv:1502.02114 [astro-ph.CO]].
  
%\cite{Komatsu:2010fb}
\bibitem{Komatsu:2010fb} 
  E.~Komatsu {\it et al.} [WMAP Collaboration], Seven-Year Wilkinson Microwave Anisotropy Probe (WMAP) observations: Cosmological interpretation, Astrophys.\ J.\ Suppl.\ Ser.  {\bf 192}, 18 (2011), [arXiv:1001.4538 [astro-ph.CO]]..

  %\cite{Aubourg:2014yra}
\bibitem{Aubourg:2014yra} 
  E.~Aubourg {\it et al.}, Cosmological implications of baryon acoustic oscillation measurements,
  Phys.\ Rev.\ D {\bf 92}, 123516 (2015),  [arXiv:1411.1074 [astro-ph.CO]].

  
  \bibitem{Ade:2015xua} 
  P.~A.~R.~Ade {\it et al.} [Planck Collaboration], Planck 2015 results. XIII. Cosmological parameters, Astron.\ Astrophys.\  {\bf 594}, A13 (2016),  [arXiv:1502.01589 [astro-ph.CO]]..

 %\cite{Sahni:1999gb}
\bibitem{Sahni:1999gb} 
  V.~Sahni and A.~A.~Starobinsky, The case for a positive cosmological $\Lambda$-term, Int.\ J.\ Mod.\ Phys.\ D {\bf 9}, 373 (2000),  [astro-ph/9904398].
    
%\cite{Peebles:2002gy}
\bibitem{Peebles:2002gy} 
  P.~J.~E.~Peebles and B.~Ratra, The cosmological constant and dark energy, Rev.\ Mod.\ Phys.\  {\bf 75}, 559 (2003),  [astro-ph/0207347].
  
  %\cite{Copeland:2006wr}
\bibitem{Copeland:2006wr} 
  E.~J.~Copeland, M.~Sami and S.~Tsujikawa, Dynamics of dark energy, Int.\ J.\ Mod.\ Phys.\ D {\bf 15}, 1753 (2006),  [hep-th/0603057].

  %\cite{Bamba:2012cp}
\bibitem{Bamba:2012cp} 
  K.~Bamba, S.~Capozziello, S.~Nojiri and S.~D.~Odintsov, Dark energy cosmology: The equivalent description via different theoretical models and cosmography tests, Astrophys.\ Space Sci.\  {\bf 342}, 155 (2012),  [arXiv:1205.3421 [gr-qc]].

%\cite{Weinberg:1988cp}
\bibitem{Weinberg:1988cp} 
  S.~Weinberg, The cosmological constant problem, Rev.\ Mod.\ Phys.\  {\bf 61}, 1 (1989).

  
    %\cite{Carroll:2000fy}
\bibitem{Carroll:2000fy} 
  S.~M.~Carroll, The cosmological constant, Living Rev.\ Relativity \  {\bf 4}, 1 (2001), [astro-ph/0004075].

  
    %\cite{Sahni:2004ai}
\bibitem{Sahni:2004ai} 
  V.~Sahni, Dark matter and dark energy, Lect.\ Notes Phys.\  {\bf 653}, 141 (2004), [astro-ph/0403324]..


  
  %\cite{Bull:2015stt}
\bibitem{Bull:2015stt} 
  P.~Bull {\it et al.}, Beyond $\Lambda$CDM: Problems, solutions, and the road ahead, Phys.\ Dark Univ.\  {\bf 12}, 56 (2016), [arXiv:1512.05356 [astro-ph.CO]].


%\cite{Zhao:2017cud}
\bibitem{Zhao:2017cud} 
  G.~B.~Zhao {\it et al.}, Dynamical dark energy in light of the latest observations, Nature \ Astron.\  {\bf 1}, 627 (2017),  [arXiv:1701.08165 [astro-ph.CO]]..
  
  %\cite{Clifton:2011jh}
\bibitem{Clifton:2011jh} 
  T.~Clifton, P.~G.~Ferreira, A.~Padilla and C.~Skordis, Modified gravity and cosmology, Phys.\ Rep.\  {\bf 513}, 1 (2012), [arXiv:1106.2476 [astro-ph.CO]]..

  %\cite{DeFelice:2010aj}
\bibitem{DeFelice:2010aj} 
  A.~De Felice and S.~Tsujikawa, $f(R)$ theories, Living Rev.\ Relativity \  {\bf 13}, 3 (2010), [arXiv:1002.4928 [gr-qc]].
  
  %\cite{Capozziello:2011et}
\bibitem{Capozziello:2011et} 
  S.~Capozziello and M.~De Laurentis, Extended theories of gravity, Phys.\ Rep.\  {\bf 509}, 167 (2011), [arXiv:1108.6266 [gr-qc]].
 
  
%\cite{Nojiri:2017ncd}
\bibitem{Nojiri:2017ncd} 
  S.~Nojiri, S.~D.~Odintsov and V.~K.~Oikonomou, Modified gravity theories on a nutshell: Inflation, bounce and late-time evolution, Phys.\ Rep.\  {\bf 692}, 1 (2017)
  %doi:10.1016/j.physrep.2017.06.001
  [arXiv:1705.11098 [gr-qc]].

%\cite{Nojiri:2010wj}
\bibitem{Nojiri:2010wj} 
  S.~Nojiri and S.~D.~Odintsov, Unified cosmic history in modified gravity: From $F(R)$ theory to Lorentz non-invariant models, Phys.\ Rep.\  {\bf 505}, 59 (2011), [arXiv:1011.0544 [gr-qc]].
 
  %\cite{Harko:2010mv}
\bibitem{Harko:2010mv} 
  T.~Harko and F.~S.~N.~Lobo, $f(R, \mathcal{L}_{\rm m}$) gravity, Eur.\ Phys.\ J.\ C {\bf 70}, 373 (2010), [arXiv:1008.4193 [gr-qc]].

  
 %\cite{Harko:2010vs}
\bibitem{Harko:2010vs} 
  T.~Harko, Galactic rotation curves in modified gravity with nonminimal coupling between matter and geometry, Phys.\ Rev.\ D {\bf 81}, 084050 (2010), [arXiv:1004.0576 [gr-qc]].

  
  %\cite{Poplawski:2006ey}
\bibitem{Poplawski:2006ey} 
  N.~J.~Poplawski, A Lagrangian description of interacting dark energy, arXiv: gr-qc/0608031.
  %%CITATION = GR-QC/0608031;%%
  %49 citations counted in INSPIRE as of 01 Apr 2017
  
  \bibitem{Harko:2011kv} 
  T.~Harko, F.~S.~N.~Lobo, S.~Nojiri and S.~D.~Odintsov, $f(R,T)$ gravity, Phys.\ Rev.\ D {\bf 84}, 024020 (2011), [arXiv:1104.2669 [gr-qc]].

%\cite{Moraes:2016gpe}
\bibitem{Moraes:2016gpe} 
  P.~H.~R.~S.~Moraes and J.~R.~L.~Santos, A complete cosmological scenario from $f(R,T^{\phi })$ gravity theory, Eur.\ Phys.\ J.\ C {\bf 76}, 60 (2016), [arXiv:1601.02811 [gr-qc]].


%\cite{Odintsov:2013iba}
\bibitem{Odintsov:2013iba} 
  S.~D.~Odintsov and D.~Sáez-Gómez, $f(R, T, R_{\mu\nu} T^{\mu\nu})$ gravity phenomenology and $\Lambda$CDM universe, Phys.\ Lett.\ B {\bf 725}, 437 (2013), [arXiv:1304.5411 [gr-qc]].


%\cite{Haghani:2013oma}
\bibitem{Haghani:2013oma} 
  Z.~Haghani, T.~Harko, F.~S.~N.~Lobo, H.~R.~Sepangi, and S.~Shahidi, Further matters in space-time geometry: $f(R,T,R_{\mu\nu}T^{\mu\nu})$ gravity, Phys.\ Rev.\ D {\bf 88}, 044023 (2013), [arXiv:1304.5957 [gr-qc]].

  %\cite{Haghani:2014ina}
\bibitem{Haghani:2014ina} 
  Z.~Haghani, T.~Harko, H.~R.~Sepangi, and S.~Shahidi, Matter may matter, Int.\ J.\ Mod.\ Phys.\ D {\bf 23}, 12  (2014), [arXiv:1304.5957 [gr-qc]].

 
%\cite{Arik:2013sti}
\bibitem{Arik:2013sti} 
  N.~Kat{\i}rc{\i} and M.~Kavuk, $ f(R,T_{\mu\nu}T^{\mu\nu})$ gravity and Cardassian-like expansion as one of its consequences, Eur.\ Phys.\ J.\ Plus {\bf 129}, 163 (2014), [arXiv:1302.4300 [gr-qc]].

    %\cite{Harko:2014sja}
\bibitem{Harko:2014sja} 
  T.~Harko, F.~S.~N.~Lobo, G.~Otalora and E.~N.~Saridakis, Nonminimal torsion-matter coupling extension of $f(T)$ gravity, Phys.\ Rev.\ D {\bf 89}, 124036 (2014), [arXiv:1404.6212 [gr-qc]].

  
      %\cite{Harko:2014aja}
\bibitem{Harko:2014aja} 
  T.~Harko, F.~S.~N.~Lobo, G.~Otalora and E.~N.~Saridakis, $f(T,\mathcal{T})$ gravity and cosmology, J. Cosmol. Astropart. Phys. 12 (2014) 021, [arXiv:1405.0519 [gr-qc]].


   \bibitem{Ashtekar:2006wn}   
 A.~Ashtekar, T.~Pawlowski and P.~Singh, Quantum nature of the big bang: Improved dynamics, Phys.\ Rev.\ D {\bf 74}, 084003 (2006),  [gr-qc/0607039].

\bibitem{Ashtekar:2011ni} 
A.~Ashtekar and P.~Singh, Loop quantum cosmology: A status report, Classical Quantum Gravity  {\bf 28}, 213001 (2011), [arXiv:1108.0893 [gr-qc]]..

\bibitem{Brax:2003fv} 
  P.~Brax and C.~van de Bruck, Cosmology and brane worlds: A review, Classical Quantum Gravity\  {\bf 20}, R201 (2003), [hep-th/0303095].

 
 \bibitem{Roshan:2016mbt} 
  M.~Roshan and F.~Shojai, Energy-momentum squared gravity, Phys.\ Rev.\ D {\bf 94}, 4 044002 (2016), [arXiv:1607.06049 [gr-qc]].

    %\cite{Board:2017ign}
\bibitem{Board:2017ign} 
  C.~V.~R.~Board and J.~D.~Barrow, Cosmological models in energy-momentum squared gravity, Phys. Rev. D \textbf{96}, 123517 (2017), arXiv:1709.09501 [gr-qc].
  
%\cite{Bertolami:2008ab}
\bibitem{Bertolami:2008ab} 
  O.~Bertolami, F.~S.~N.~Lobo and J.~Paramos, Non-minimum coupling of perfect fluids to curvature, Phys.\ Rev.\ D {\bf 78}, 064036 (2008), [arXiv:0806.4434 [gr-qc]].

  
%\cite{Faraoni:2009rk}
\bibitem{Faraoni:2009rk} 
  V.~Faraoni, The Lagrangian description of perfect fluids and modified gravity with an extra force, Phys.\ Rev.\ D {\bf 80}, 124040 (2009), [arXiv:0912.1249 [astro-ph.GA]].

  %\cite{Uzan:2010pm}
\bibitem{Uzan:2010pm} 
  J.~P.~Uzan, Varying constants, gravitation and cosmology, Living Rev.\ Relativity \  {\bf 14}, 2 (2011), [arXiv:1009.5514 [astro-ph.CO]]..


  %\cite{Freese:2002sq}
\bibitem{Freese:2002sq} 
  K.~Freese and M.~Lewis, Cardassian expansion: A model in which the universe is flat, matter dominated, and accelerating, Phys.\ Lett.\ B {\bf 540}, 1 (2002), [astro-ph/0201229].

  
    %\cite{Chung:1999zs}
\bibitem{Chung:1999zs} 
  D.~J.~H.~Chung and K.~Freese, Cosmological challenges in theories with extra dimensions and remarks on the horizon problem, Phys.\ Rev.\ D {\bf 61}, 023511 (2000), [hep-ph/9906542].

 
    %\cite{Kamenshchik:2001cp}
\bibitem{Kamenshchik:2001cp} 
  A.~Y.~Kamenshchik, U.~Moschella and V.~Pasquier, An alternative to quintessence, Phys.\ Lett.\ B {\bf 511}, 265 (2001),  [gr-qc/0103004].


  %\cite{Bento:2002ps}
\bibitem{Bento:2002ps} 
  M.~C.~Bento, O.~Bertolami and A.~A.~Sen, Generalized Chaplygin gas, accelerated expansion and dark energy matter unification, Phys.\ Rev.\ D {\bf 66}, 043507 (2002),  [gr-qc/0202064].

  
  %\cite{Ananda:2005xp}
\bibitem{Ananda:2005xp} 
  K.~N.~Ananda and M.~Bruni, Cosmo-dynamics and dark energy with non-linear equation of state: a quadratic model, Phys.\ Rev.\ D {\bf 74}, 023523 (2006),  [astro-ph/0512224].
   
   %\cite{Josset:2016vrq}
\bibitem{Josset:2016vrq} 
  T.~Josset, A.~Perez and D.~Sudarsky, Dark energy from violation of energy conservation, Phys.\ Rev.\ Lett.\  {\bf 118}, 021102 (2017), [arXiv:1604.04183 [gr-qc]].

\bibitem{Shabani:2017vns} 
  H.~Shabani and A.~H.~Ziaie, Consequences of energy conservation violation: Late-time solutions of $\Lambda (\mathsf{T}) \mathsf{CDM}$ subclass of $f(\mathsf{R},\mathsf{T})$ gravity using dynamical system approach, Eur.\ Phys.\ J.\ C {\bf 77}, 282 (2017), [arXiv:1702.07380 [gr-qc]].

 %\cite{Lovelock:1971yv}
\bibitem{Lovelock:1971yv} 
  D.~Lovelock, The Einstein tensor and its generalizations,  J.\ Math.\ Phys.\  {\bf 12}, 498 (1971).

\bibitem{Lovelock:1972vz} 
  D.~Lovelock, The four-dimensionality of space and the Einstein tensor, J.\ Math.\ Phys.\  {\bf 13}, 874 (1972).

\bibitem{ACEK} 
O.~Akarsu, J.~D.~Barrow, S.~\c C{\i}k{\i}nto\u glu, K.~Y.~Ek\c{s}i and N.~Katirci, Constraints on energy-momentum squared gravity from neutron stars and its cosmological implications, arXiv:1802.02093.

\bibitem{Straumann}
N.~Straumann, General Relativity: With Applications to Astrophysics. (Springer-Verlag, 2004).
  
%\cite{Barrow:1976rda}
\bibitem{Barrow:1976rda} 
  J.~Barrow, Light elements and the isotropy of the universe, Mon.\ Not.\ R.\ Astron.\ Soc.\  {\bf 175}, 359 (1976).



%\cite{Rubano:2001xi}
\bibitem{Rubano:2001xi} 
  C.~Rubano and J.~D.~Barrow, Scaling solutions and reconstruction of scalar field potentials, Phys.\ Rev.\ D {\bf 64}, 127301 (2001),  [gr-qc/0105037].

%\cite{Caldwell:1999ew}
\bibitem{Caldwell:1999ew}
  R.~R.~Caldwell, A phantom menace?, Phys.\ Lett.\ B {\bf 545}, 23 (2002), [astro-ph/9908168].

  
  %\cite{Caldwell:2003vq}
\bibitem{Caldwell:2003vq} 
  R.~R.~Caldwell, M.~Kamionkowski and N.~N.~Weinberg, Phantom energy and cosmic doomsday, Phys.\ Rev.\ Lett.\  {\bf 91}, 071301 (2003), [astro-ph/0302506].


 \bibitem{carroll03}
S.~M.~Carroll, M.~Hoffman, and M.~Trodden, Can the dark energy equation-of-state parameter w be less than $-1$?, Phys. Rev. D {\bf 68}, 023509 (2003), [arXiv:astro-ph/0301273].

\bibitem{cline04}
J.~M.~Cline, S.~Jeon, and G.~D.~Moore, The phantom menaced: Constraints on low-energy effective ghosts, Phys. Rev. D {\bf 70}, 043543, (2004), [arXiv:hep-ph/0311312].

  %\cite{Farooq:2013hq}
\bibitem{Farooq:2013hq} 
  O.~Farooq and B.~Ratra, Hubble parameter measurement constraints on the cosmological deceleration-acceleration transition redshift, Astrophys.\ J.\  {\bf 766}, L7 (2013),  [arXiv:1301.5243 [astro-ph.CO]].

   %\cite{Chen:2016uno}
\bibitem{Chen:2016uno} 
  Y.~Chen, S.~Kumar and B.~Ratra, Determining the Hubble constant from Hubble parameter measurements, Astrophys.\ J.\  {\bf 835}, 1 86 (2017), [arXiv:1606.07316 [astro-ph.CO]].

  
  %\cite{Lewis:2002ah}
\bibitem{Lewis:2002ah} 
  A.~Lewis and S.~Bridle, Cosmological parameters from CMB and other data: A Monte Carlo approach,  Phys.\ Rev.\ D {\bf 66}, 103511 (2002), [astro-ph/0205436].

 
 %\cite{Simon:2004tf}
\bibitem{Simon:2004tf} 
  J.~Simon, L.~Verde and R.~Jimenez, Constraints on the redshift dependence of the dark energy potential, Phys.\ Rev.\ D {\bf 71}, 123001 (2005), [astro-ph/0412269].

  
  \bibitem[Stern et~al.(2010)]{Sternetal2010} {D.~Stern}, {\textit{et}~al.}, Cosmic chronometers: Constraining the equation-of-state of dark energy. II. A spectroscopic catalog of red galaxies in galaxy clusters, J. Cosmol. Astropart. Phys. 02 (2010) 008, [arXiv:0907.3152].
         
    \bibitem[{Moresco} et~al.(2012)]{Morescoetal2012}
     {M. Moresco} {\textit{et}~al.}, Improved constraints on the expansion rate of the Universe up to $z\sim1.1$ from the spectroscopic evolution of cosmic chronometers, J. Cosmol. Astropart. Phys. 08, (2012) 006, [arXiv:1201.3609].


\bibitem[{{Busca} {et~al.}(2013)}]{Buscaetal2013}
{N. G. Busca} {\textit{et}~al.}, Baryon acoustic oscillations in the Ly-$\alpha$ forest of BOSS quasars, Adyton. Astrophys. {\bf 552}, A96 (2013), [arXiv:1211.2616].

\bibitem[{{Zhang} {et~al.}(2014)}]{Zhangetal2014}
        {C. Zhang} {\textit{et}~al.}, Four New Observational $H(z)$ Data From Luminous Red Galaxies of Sloan Digital Sky Survey Data Release Seven, Res. Astron. Astrophys., {\bf 14}, 1221 (2014), [arXiv:1207.4541].
        
  \bibitem[{{Zhang} {et~al.}(2014)}]{Blakeetal2012}
        {C. Blake} {\textit{et}~al.}, The WiggleZ Dark Energy Survey: Joint measurements of the expansion and growth history at $z <1$, Mon. Not. R. Astron. Soc. {\bf 425}, 405 (2012), [arXiv:1204.3674].
  
         \bibitem[{{Chuang} \& {Wang}(2013)}]{ChuangWang2013}
         {C.-H. Chuang} and {Y. Wang}, Modeling the anisotropic two-point galaxy correlation function on small scales and single-probe measurements of $H(z)$, $D_A(z)$, and $f(z)\sigma_8(z)$ from the Sloan Digital Sky Survey DR7 luminous red galaxies", Mon. Not. R. Astron. Soc., {\bf 435}, 255 (2013), [arXiv:1209.0210].

  %\cite{Chavanis:2016pcp}
\bibitem{Chavanis:2016pcp} 
  P.~H.~Chavanis and S.~Kumar, Comparison between the Logotropic and $\Lambda$CDM models at the cosmological scale, J. Cosmol. Astropart. Phys. 05 (2017) 018, [arXiv:1612.01081 [astro-ph.CO]].

\bibitem{Li:2007xw} 
  B.~Li, J.~D.~Barrow and D.~F.~Mota, The cosmology of Ricci-tensor-squared gravity in the Palatini variational approach, Phys.\ Rev.\ D {\bf 76}, 104047 (2007), [arXiv:0707.2664 [gr-qc]].

%\cite{Horndeski:1974wa}
\bibitem{Horndeski:1974wa} 
  G.~W.~Horndeski, Second-order scalar-tensor field equations in a four-dimensional space, Int.\ J.\ Theor.\ Phys.\  {\bf 10}, 363 (1974).
  

%\cite{Moraes:2016zdw}
\bibitem{Moraes:2016zdw} 
  P.~H.~R.~S.~Moraes, R.~A.~C.~Correa and G.~Ribeiro, The Starobinsky model within the $f(R,T)$ formalism as a cosmological model, arXiv:1701.01027, arXiv:1701.01027 [gr-qc].


\end{thebibliography}
\end{document}